\documentclass[]{svmult}
\usepackage{natbib}
\usepackage{makeidx}         
\usepackage{graphicx}
\usepackage{amsfonts}
\usepackage{amsmath}
\usepackage{lineno}

\def\En{\mathcal{E}}
\def\S{\mathcal{S}}
\def\en{E}
\def\E{{\mathbb E}} 
\def\ZZ{{\mathbb Z}}       

\def\R{\mathbb R}
\def\P{\mathbb P}

\def\V{\mbox{Var}}
\def\S{\mathcal S}

\makeindex

\spnewtheorem{hypothesis}[theorem]{Hypothesis}{\bf}{\it}
\begin{document}

\bibliographystyle{plainnat}

\title*{Coexistence in the face of uncertainty}
\author{Sebastian J. Schreiber}
\authorrunning{S.J. Schreiber}
\institute{Sebastian J. Schreiber \at Department of Evolution and Ecology, University of California, Davis, CA 95616 USA \email{sschreiber@ucdavis.edu}}

\maketitle

\vskip -2in
\begin{quote} Lest men believe your tale untrue, keep probability in view. --John Gay\end{quote} 

\begin{abstract}{ Over the past century, nonlinear difference and differential equations have been used to understand conditions for coexistence of interacting populations. However, these models fail to account for random fluctuations due to demographic and environmental stochasticity which are experienced by all populations. I review some recent mathematical results about persistence and coexistence for models accounting for each of these forms of stochasticity. Demographic stochasticity stems from populations and communities consisting of a finite number of interacting individuals, and often are represented by Markovian models with a countable number of states. For closed populations in a bounded world, extinction occurs in finite time but may be preceded by long-term transients. Quasi-stationary distributions (QSDs) of these Makov models characterize this meta-stable behavior. For sufficiently large ``habitat sizes'', QSDs are shown to concentrate on the positive attractors of deterministic models. Moreover, the probability extinction decreases exponentially with habitat size. Alternatively, environmental stochasticity stems from fluctuations in environmental conditions which influence survival, growth, and reproduction. Stochastic difference equations can be used to model the effects of environmental stochasticity on population and community dynamics. For these models, stochastic persistence corresponds to empirical measures placing arbitrarily little weight on arbitrarily low population densities. Sufficient and necessary conditions for stochastic persistence are reviewed. These conditions involve weighted combinations of Lyapunov exponents corresponding to ``average'' per-capita growth rates of rare species. The results are illustrated with how climatic variability influenced the dynamics of Bay checkerspot butterflies, the persistence of coupled sink populations, coexistence of competitors through the storage effect, and stochastic rock-paper-scissor communities. Open problems and conjectures are presented.} 
\end{abstract}

\keywords{Random difference equations, stochastic population dynamics, coexistence, quasi-stationary distributions, demographic noise, environmental stochasticity, Markov chains}

\vskip 0.1in
{\large To appear as a refereed chapter in \emph{Recent Progress and Modern Challenges in Applied Mathematics, Modeling and Computational Science} in \emph{Fields Institute Communication Series} edited by Roderick Melnik, Roman Makarov, and Jacques Belair}

\section{Introduction}
A long standing, fundamental question in biology is ``what are the minimal conditions to ensure the long-term persistence of a population or the long-term coexistence of interacting species?'' The answers to this question are essential for guiding conservation efforts for threatened and endangered species, and identifying mechanisms that maintain biodiversity. Mathematical models have and continue to play an important role in identifying these potential mechanisms and, when coupled with empirical work, can test whether or not a given mechanism is operating in a specific population or ecological community~\citep{adler-etal-10}.  Since the pioneering work  of \citet{Lotka-25} and \citet{volterra-26} on competitive and predator--prey interactions, \citet{thompson-24}, \citet{nicholson-bailey-35} on host--parasite interactions, and \citet{kermack-mckendrick-27} on disease outbreaks, nonlinear difference and differential equations have been used to understand conditions for persistence of populations or communities of interacting species. For these deterministic models,  persistence or species coexistence is often equated with an attractor bounded away from the extinction states in which case persistence holds over an infinite time horizon~\citep{jtb-06}. However (with apologies to John Gay), lest biologists believe this theory untrue, the models need to keep probability in view. That is, all natural populations exhibit random fluctuations due to mixture of intrinsic and extrinsic factors known as demographic and environmental stochasticity. The goal of this chapter is to present models that account for these random fluctuations, review some mathematical methods for analyzing these stochastic models, and illustrate how these random fluctuations hamper or facilitate population persistence and species coexistence.  \index{species coexistence}\index{population persistence}

Demographic stochasticity corresponds to random fluctuations due to populations consisting of a finite number of individuals whose fates aren't perfectly correlated. \index{demographic stochasticity} That is, even if all individuals in a population appear to be identical, some undetectable differences between individuals (e.g. in their physiology or microenvironment) result in some individuals dying while others survive. To capture these ``unknowable'' differences, models can assign the same probabilities of dying to each individuals and treat survival amongst individuals as independent flips of a coin -- heads life, tails death. Similarly, surviving individuals may differ in the number of offspring they produce despite appearing to be identical. To capture these unknowable differences,  the number of offspring produced by these individuals are modeled as independent draws from the same probability distribution. The resulting stochastic models accounting for these random fluctuations typically correspond to Markov chains on a finite or countable state space\footnote{See, however, the discussion for biologically motivated uncountable state spaces.} e.g. the numbers of individuals,  $0,1,2,3,\dots$, in a population. When these models represent populations or communities whose numbers tend to stay bounded and have no immigration, the populations in these models always go extinct in finite time~\citep{chesson-78}. Hence, unlike deterministic models, the asymptotic behavior of these stochastic models is trivial: eventually no one is left. This raises the following basic question about the relationship between models accounting for demographic stochasticity and their deterministic counterparts:  
\begin{quote}
``Any population allowing individual variation in reproduction, ultimately dies out--unless it grows beyond all limits, an impossibility in a bounded world. Deterministic population mathematics on the contrary allows stable asymptotics. Are these artifacts or do they tell us something interesting about quasi-stationary stages of real or stochastic populations?'' -- Peter \citet{jagers-10}
\end{quote}
As it turns out, there is a strong correspondence between the quasi-stationary behavior of the stochastic models and the attractors of an appropriately defined mean-field model. Moreover, this correspondence highlights a universal scaling relationship between extinction times and the size of the habitat where the species live.  These results and their applications are  the focus of the first half of this review. 

While demographic stochasticity affects individuals independently, environmental stochasticity concerns correlated  demographic responses (e.g. increased survival, growth or reproduction) among individuals.   \index{demographic stochasticity}\index{environmental stochasticity}These correlations often stem from individuals experiencing  similar fluctuations in environmental conditions (e.g. temperature, precipitation, winds) which impact their survival, growth, or reproduction. Models driven by randomly fluctuating parameters or brownian motions, such as random difference equations or stochastic differential equations, can capture these sources of random fluctuations. Unlike models for demographic stochasticity, these Markov chains always live on uncountable state spaces where the non-negative reals represent densities of populations of sufficiently large size that one can ignore the effects of being discrete and finite. Consequently, like their deterministic counterparts, extinction in these random difference equations only occurs asymptotically, and persistence is equated with tendency to stay away from low densities~\citep{chesson-82}. Understanding what this exactly means, reviewing methods for verifying this stochastic form of persistence, and applying these methods to gain insights about population persistence and species coexistence are the focus of the second half of this review.  

Of course, all population systems experience a mixture of demographic and environmental stochasticity. While the theoretical biology literature is replete with models accounting for each of these forms of noise separately, I know of no studies that rigorously blend the results presented in this review. Hence, I conclude by discussing some open problems and future challenges at this mathematical interface. 

\section{Demographic stochasticity}

To model finite populations and account for demographic stochasticity, we consider Markov chains on a countable state space which usually is the non-negative cone of the integer lattice. Many of these stochastic models have a deterministic counterpart, sometimes called the ``deterministic skeleton'' or the ``mean field model''. As I discuss below, these deterministic models can provide some useful insights about the transient behavior of the stochastic models and when coupled with large deviation theory provide insights into the length of these transients. 

To get a flavor of the types of models being considered, lets begin with a stochastic counterpart to the discrete-time Lotka-Volterra equations. This example motivates the main results and will illustrate their applicability.

\begin{example}[Poisson Lotka-Volterra Processes]~\label{sec:LV}  \index{Lotka-Volterra Processes} The continuous time Lotka-Volterra equations form the bedrock for much of community ecology theory. While there are various formulations of their discrete-time counterparts, a particularly pleasing one that retains several key dynamical features of the continuous-time models was studied by \citet{hofbauer-etal-87}. These models  keep track of the densities $x_t=(x_{1,t},\dots,x_{k,t})$ of $k$ interacting species, where the subscripts denote the species identity $i$ and time $t$ (e.g. year or day). As with the classical continuous time equations, there is a matrix $A=(a_{ij})_{i,j}$ where $a_{ij}$ corresponds to the ``per-capita'' effect of species $j$ on species $i$ and a vector $r=(r_1,\dots,r_k)$ of the ``intrinsic per-capita growth rates'' for all of the species. With this notation, the equations take on the form:
\begin{equation}\label{eq:LVd}
x_{i,t+1} = x_{i,t} \exp\left(r_i + \sum_j a_{ij} x_{j,t}\right)=:F_i(x_t) \mbox{ with } i=1,2,\dots,k.
\end{equation}
The state space for these dynamics are given by the non-negative orthant \[\R_+^k=\{x\in \R^k: x_i \ge 0\mbox{ for all }i\}\] of the $k$-dimensional Euclidean space $\R^k$.

To define the Poisson Lotka-Volterra process, let $1/\varepsilon$ be the size of the habitat in which the species live. Let $N_t^\varepsilon=(N_{1,t}^\varepsilon,\dots,N_{k,t}^\varepsilon)$ denote the vector of species abundances which are integer-valued.   Then the density of species $i$ is $X_{i,t}^\varepsilon=\varepsilon N_{i,t}^\varepsilon$. Over the next time step, each individual replaces itself with a Poisson number of individuals with mean 
\[
\exp\left(r_i + \sum_j a_{ij} X_{j,t}^\varepsilon\right).
\]
If the individuals update independent of one another, then  $N_{i,t+1}^\varepsilon$ is a sum of $N_{i,t}^\varepsilon$ independent Poisson random variables. Thus, $N_{i,t+1}^\varepsilon$ is also Poisson distributed with mean 
\[
N_{i,t}^\varepsilon\exp\left(r_i + \sum_j a_{ij} X_{j,t}^\varepsilon \right)=F_i(X_t^\varepsilon)/\varepsilon.
\]
Namely, 
 \begin{equation}\label{eq:poisson}
  \P[X_{i,t+1}^\varepsilon= \varepsilon j | X_t^\varepsilon=x]=\P[N_{i,t+1}^\varepsilon = j | X_t^\varepsilon=x]=\exp(-F_i(x)/\varepsilon) \frac{(F_i(x)/\varepsilon)^j}{j!}. 
 \end{equation}
 The state space for $N_t^\varepsilon$ is the non-negative, $k$ dimensional integer lattice 
 \[ \ZZ^k_+=\{(z_1,\dots,z_k): z_i\mbox{ are non-negative integers}\}
 \]
 while the state space for $X_t^\varepsilon$ is the non-negative, rescaled integer lattice
 \[
 \varepsilon \ZZ^k_+ = \{  (\varepsilon z_1,\dots,\varepsilon z_k): z_i\mbox{ are non-negative integers}\}.
 \]

 \begin{figure}
 \begin{center}
 \includegraphics[width=6in]{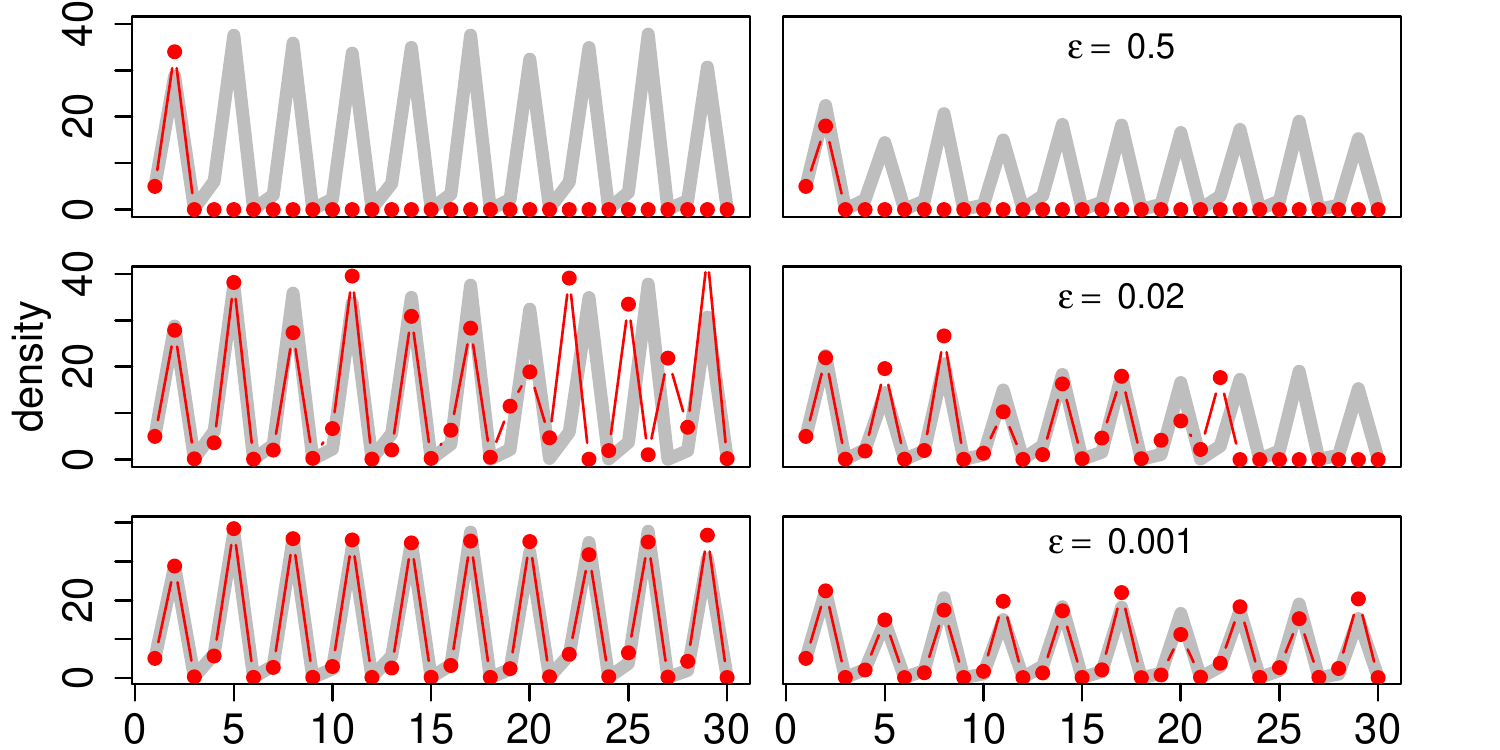}
 \end{center}
 \caption{Realizations of a Poisson Lotka-Voltera process with two competing species (species $1$ on the left, species $2$ on the right). The deterministic dynamics are shown as a thick gray line.  Stochastic realizations are shown in red. Each row corresponds to a different habitat size $1/\varepsilon$. Parameter values: $A$ is the matrix with rows $(-0.2,0.1)$, $(-0.15,0.2)$, and $r=(3.25,3.25)$ for the model described in Example~\ref{sec:LV}.
}\label{fig:1}
 \end{figure}

Now consider a solution to deterministic model $x_t$  and the stochastic process $X_t^\varepsilon$  initiated at the same densities $x_0=X_0^\varepsilon =x$. To see how likely $X_t^\varepsilon$  deviates from $x_t$, we use Chebyshev's inequality.  As the mean and variance of a Poisson random variable are equal, Chebyshev's inequality implies
 \begin{equation}\label{cheb}
 \P\left[ |X_{i,1}^\varepsilon-x_{i,1}|\ge \delta \Big| X_0^\varepsilon=x_0=x \right]\le \frac{\V[X_{i,1}^\varepsilon]}{\delta^2}= \frac{\varepsilon^2 \V[N_{i,1}^\varepsilon]}{\delta^2}= \frac{\varepsilon F_i(x)}{\delta^2}
 \end{equation}
 where $\V[X]$ denotes the variance of a random variable $X$.  In words, provided the habitat size $1/\varepsilon$ is sufficiently large, a substantial deviation between $X_1^\varepsilon$ and $x_1$ is unlikely. In fact, one can show that over any finite time interval $[1,T]$, the stochastic dynamics are likely to be close to the deterministic dynamics over the time interval $[1,T]$ provided the  habitat size $1/\varepsilon$ is sufficiently large:
 \begin{equation}\label{approx}
 \lim_{\varepsilon\to 0}\P\left[\max_{1\le i\le k,1\le t\le T} |X_{i,t}^\varepsilon-x_{i,t}|\ge \delta \Big| X_0^\varepsilon=x_0=x \right]=0.
 \end{equation}
\noindent Figure~\ref{fig:1} illustrates this fact for a Poisson Lotka-Volterra process with two competing species. Equation~\eqref{approx} is the discrete-time analog of a result derived by \citet{kurtz-81} for continuous-time Markov chains. \citet{kurtz-81} also provides ``second-order'' approximations for finite time  intervals using Gaussian processes and stochastic differential equations. While these approximations are also useful for discrete-time models,  we do not review them here. 

Despite $X_t^\varepsilon$ stochastically tracking $x_t$ with high probability for long periods of time, eventually their behavior diverges as Poisson Lotka-Volterra processes go extinct in finite time or exhibit unbounded growth. 

 \begin{proposition}\label{prop:extinct} Let $X_t^\varepsilon$ be a Poisson Lotka-Volterra process with $\varepsilon>0$. Then 
 \[
 \P\left[ \{ X_t ^\varepsilon=0 \mbox{ for some }t\} \cup \{ \lim_{t\to\infty} \sum_i X_{i,t}^\varepsilon=\infty\}\right]=1
 \]
 Furthermore, if $F$ is pre-compact i.e. $F(\R^k_+)\subset [0,m]^k$ for some $m\ge 0$, then 
 \[
 \P\left[ \{ X_t^\varepsilon =0 \mbox{ for some }t\}\right]=1
 \]
 \end{proposition}

The strategy used to prove the first statement of the proposition is applicable to many models of closed populations. The key ingredients are that there is a uniform lower bound to the probability of any individual dying, and  individuals die independently of one another~\citep{chesson-78}. Proving, however, that extinction always occurs with probability one  requires additional elements which aren't meet by all ecological models, but is meet for ``realistic'' models. 
 
 \newpage
 \begin{proof}
 For the first assertion, take any integer $m>0$. Let 
 \[
 \beta=\min_{x\in [0,m]^k} \P[X_1^\varepsilon=0|X_0^\varepsilon=x]=\min_{x\in [0,m]^k } \exp\left(-\sum_{i=1}^k F_i(x)/\varepsilon\right)>0.
 \] 
Next we use the following standard result in Markov chain theory~\citep[Theorem 2.3 in Chapter 5]{durrett-96}.

\begin{proposition}\label{thm:mc}
Let $X$ be a Markov chain and suppose that
$$ \P\left[\bigcup_{s=1}^{\infty} \{ X_{t+s}\in C\}\Big|X_t\right] \ge \beta>0\mbox{ on } \{X_t\in B\}. $$
Then
$$P\left[\{X_t\mbox{ enters } B \mbox{ infinitely often}\}\setminus \{X_t\mbox{ enters } C\mbox{ infinitely often}\}\right]=0. $$
\end{proposition}

Let $\mathcal{B}_m=\{X_t^\varepsilon \mbox{ enters } [0,m]^k \mbox{ infinitely often}\}$ and $\mathcal{E}=\{X_t ^\varepsilon=0 \mbox{ for some }t\}$. Proposition~\ref{thm:mc} with $B=[0,m]^k$ and $C=\{0\}$ implies that 
\begin{equation}\label{io}
\P\left[\mathcal{B}_m\setminus \mathcal {E}\right]=0.
\end{equation} The complement of the event $\cup_{m} \mathcal {B}_m$ equals the event $\mathcal{A}=\{ \lim_{t\to\infty} \sum_i X_{i,t}^\varepsilon=\infty\}$. As $\mathcal{B}_m$ is an increasing sequence of events, 
\begin{eqnarray*}
1&=& \P\left[ \mathcal{A}\cup \{\cup_m \mathcal {B}_m\}\right]\\
&=& \lim_{m\to\infty}\P\left[ \mathcal{A}\cup \mathcal {B}_m\right]\\
&\le & \lim_{m\to\infty}\P\left[ \mathcal{A}\cup \mathcal {E}\right]
\end{eqnarray*}
where the final inequality follows from \eqref{io}. This completes the proof of the first assertion. 

To prove the second assertion, assume that there exists $m>0$ such that $F(x)\in [0,m]^k$ for all $x\in \R_+^k$ i.e. $F$ is pre-compact.  Define 
\begin{eqnarray*}
\beta&=&\inf_{x\in\R_+^k} \P[X_{t+1}^\varepsilon =0 | X_t =x]\\
&=& \inf_{x\in \R_+^k}  \exp\left(-\sum_i F_i(x)/\varepsilon\right)\\
&\ge&  \exp(-k\,m/\varepsilon)\\
\end{eqnarray*}
Applying Proposition~\ref{thm:mc} with $B=\R_+^k$ and $C=\{0\}$ completes the proof of the second assertion. \qed
 \end{proof} 
\end{example} 

Equation~\eqref{approx} and Proposition~\ref{prop:extinct} raise two fundamental questions about these stochastic, finite population models: How long before extinction occurs? Prior to extinction what can one say about the transient population dynamics? To get some insights into both of these questions, we build on the  work of \citet{freidlin-wentzell-98} and \citet{kifer-88} on random perturbations of dynamical systems, and  \citet{barbour-76} on quasi-stationary distributions.

\subsection{Random perturbations  and quasi-stationary distributions}

The Poisson Lotka-Volterra process (Example~\ref{sec:LV})  illustrates how Markovian models can be viewed as random perturbations of a deterministic model. To generalize this idea, consider a continuous, precompact\footnote{Namely,  there exists $C>0$ such that $F(\S)$ lies in $[0,C]^k$.} map $F: \S \rightarrow \S$, where $\S$ is a closed subset of $\mathbb{R}^k$. $F$ will be the deterministic skeleton of our stochastic models.  \emph{A random perturbation of $F$} is  a family of Markov chains $\{X^{\varepsilon}\}_{\varepsilon > 0}$ on $\S$ whose transition kernels \index{dynamical system!random perturbation of}
\[p^{\varepsilon}(x,\Gamma) = \mathbb{P} \left[X_{t+1}^{\varepsilon} \in \Gamma \mid X_t^{\varepsilon} = x \right] \mbox{ for all } x \in \S\mbox{ and Borel sets } \Gamma \subset \S\]
enjoy the following hypothesis:
\begin{hypothesis} \label{H} 
For any $\delta>0$, 
\[ \lim_{\varepsilon\to 0} \sup_{x \in \S} p^{\varepsilon}\left(x, \S\setminus N^{\delta}(F(x))\right)=0 \]
where $N^{\delta}(y):= \left\{x \in \S : \|y-x\|< \delta \right\}$ denotes the $\delta$-neighborhood of a point $y\in  \S$.
\end{hypothesis}
\noindent Hypothesis~\ref{H} implies that the Markov chains $X^\varepsilon$ converge to the deterministic limit as $\varepsilon \downarrow 0$ i.e. the probability  of $X_1^\varepsilon$ being arbitrarily close to $F(x)$ given $X_0^\varepsilon=x$ is arbitrarily close to one for $\varepsilon$ sufficiently small. Hence, one can view $F$ as the ``deterministic skeleton'' which gets clothed by the stochastic dynamic $X^\varepsilon$. The next example illustrates how to verify the Poisson Lotka-Volterra process is a random perturbation of the Lotka-Volterra difference equations.

\begin{example}[The Poisson Lotka-Volterra processes revisited]
Consider  the Poisson Lotka-Volterra processes from Example~\ref{sec:LV} where $F(x)=(F_1(x),\dots,F_k(x))$ and $F_i(x)=x_i\exp(r_i + \sum_j a_{ij}x_j)$ and $\S=\R^k_+$. For many natural choices of $r_i$ and $a_{ij}$, \citet{hofbauer-etal-87} have shown there exists $C>0$ such that $F(\S)\subset [0,C]^k$ i.e. $F$ is pre-compact. While the corresponding Lotka-Volterra process $X^\varepsilon$ lives on $\varepsilon\ZZ^k_+$, the process can be extended to all of $\S$ by allowing $X_0^\varepsilon$ to be any point in $\S$ and update with the transition probabilities of \eqref{eq:poisson}. With this extension, $X_1^\varepsilon$ always lies in $\varepsilon \ZZ_+^k$ and $p^\varepsilon$ is characterized by the following probabilities
\[
p^\epsilon(x,\{y\})=\prod_{i=1}^k \exp(-F_i(x)/\varepsilon) \frac{(F_i(x)/\varepsilon)^{j_i}}{{j_i}!}  \mbox{ for }y=\varepsilon (j_1,\dots,j_k)\in \varepsilon \ZZ_+^k,x\in\S
\]
and $0$ otherwise. With this extension, Hypothesis 1 for the Lotka-Volterra process follows from equation \eqref{cheb}. \end{example}

As with the Poisson Lotka-Volterra process,  stochastic models of interacting populations without immigration always have absorbing states $\S_0\subset \S$ corresponding to the loss of one or more populations. Hence, we restrict our attention to models which satisfy the following standing hypothesis:
\begin{hypothesis} \label{hy:M} The state space $\S$ can be written $\S = \S_0 \cup \S_+$, where
\begin{itemize}
\item[$\bullet$] $\S_0$ is a closed subset of $\S$;
\item[$\bullet$] $\S_0$  and $\S_+$ are positively $F$-invariant, i.e $F(\S_0) \subseteq \S_0$ and $F(\S_+) \subseteq \S_+$; 
\item[$\bullet$] the set $\S_0$ is assumed to be absorbing for the random perturbations:
\begin{equation} \label{hy:abs}  \; p^{\varepsilon}(x,\S_+)= 0, \mbox{ for all } \varepsilon >0, \, x \in \S_0.\end{equation}
\item[$\bullet$] absorption occurs in finite time with probability one:
\[
\P\left[ X_t^\varepsilon \in \S_0 \mbox{ for some } t\ge 1|X_0^\varepsilon=x\right]=1
\]
for all $x\in \S$ and $\varepsilon>0$.
\end{itemize}
\end{hypothesis}
\noindent The final bullet point implies that extinction of one or more species is inevitable in finite time. For example, Proposition~\ref{prop:extinct} implies this hypothesis for Poisson Lotka-Volterra processes whenever $F$ is pre-compact. 

Despite this eventual absorption,  the process $X^\varepsilon$ may spend exceptionally long periods of time in the set $\S_+$ of transient states provided that $\varepsilon>0$ is sufficiently small. This ``metastable'' behavior may correspond to long-term persistence of an endemic disease, long-term coexistence of interacting species as in the case of the Poisson Lotka-Volterra process, or maintenance of a genetic polymorphism. \index{species coexistence!metastability} One approach to examining these  metastable behaviors are quasi-stationary distributions which are invariant distributions when the process is conditioned on non-absorption. 
\begin{definition} \label{df:qsd}\index{quasi-stationary distribution}
A probability measure $\mu_{\varepsilon}$ on $\S_+$ is a \emph{quasi-stationary distribution (QSD)} for $p^{\varepsilon}$ provided there exists $\lambda_{\varepsilon}\in (0,1) $ such that
\[\int_{\S_+} p^{\varepsilon}(x,\Gamma) \mu_{\varepsilon}(dx)  = \lambda_{\varepsilon} \mu_{\varepsilon}(\Gamma) \mbox{ for all Borel sets } \Gamma \subset \S_+.\]
\end{definition}
\noindent Equivalently, dropping the $\varepsilon$ superscript and subscripts, a QSD $\mu$  satisfies the identity 
\[\mu(\Gamma) = \mathbb{P}_{\mu} \left[X_t \in \Gamma \mid \, \; X_t \in \S_+ \right]\mbox{ for all } t, \]
where $\mathbb{P}_{\mu}$ denotes the law of the Markov chain $\{X_t\}_{t=0}^\infty$, conditional to $X_0$ being distributed according to $\mu$. 

In the case that the Markov chain has a finite number of states and $P$ is the transition matrix (i.e. $P_{ij}=p(i,\{j\})$), \citet{darroch-seneta-65} showed that the QSD is given by $\mu(\{i\})=\pi_i$ where $\pi$ is the normalized, dominant left eigenvector of the matrix $Q$ given by removing the rows and columns of $P$ corresponding to extinction states in $\S_0$. In this case, $\lambda$ is the corresponding eigenvalue of this eigenvector. For the Poisson Lotka-Volterra processes in which the unperturbed dynamic $F$ is pre-compact, Proposition 6.1 from \citet{aap-14} implies the existence of QSDs for these processes.  Examples of these QSDs for these processes are shown in Figures~\ref{fig:QSD},~\ref{fig:Ricker}, and \ref{springer-2}. More generally, the existence of QSDs has been studied extensively by many authors  as reviewed by \citet{meleard-12}.

What do these QSD's and $\lambda$ tell us about the behavior of the stochastic process? From the perspective of metastability,  \citet{} QSDs often exhibit the following property:
\[\mu(\Gamma) = \lim_{t \rightarrow + \infty} \mathbb{P} \left[X_t \in \Gamma \mid \, X_t \in \S_+ ,X_0=x\right]\]
where the limit exists and is independent of the initial state $x \in \S_+$. In words, the QSD describes the probability distribution of $X_t$, conditioned on non-extinction, far into the future. Hence, the QSD provides a statistical description of the meta-stable behavior of the process.  The eigenvalue, $\lambda$  provides information about the length of the metastable behavior of $X_t$. Specifically, given that the process is following the QSD (e.g. $X_0$ is distributed like $\mu$),  and $\lambda$ equals the probability of persisting in the next time step. Thus, the mean time to extinction is $\frac{1}{1-\lambda}$. \citet{grimm-wissel-04} call $\frac{1}{1-\lambda}$, the ``intrinsic mean time to extinction'' and, convincingly, argue that it is a fundamental statistic for comparing extinction risk across stochastic models. \index{extinction!mean time to}\index{extinction!probability}

\subsection{Positive attractors, intrinsic extinction risk, and metastability} 

When the habitat size is sufficiently large i.e. $\varepsilon$ is small, there is a strong relationship between the existence of attractors in $\S_+$ (i.e. ``positive'' attractors) for the unperturbed system $F$ and the quasi-stationary distributions of $X^\varepsilon$. \index{attractor!positive} This relationship simultaneously provides information about the metastable behavior of the stochastic model and intrinsic probability of extinction, $1-\lambda_\varepsilon$. To make this relationship mathematically rigorous, we need to strengthen Hypotheses \ref{H} and \ref{hy:M}. \citet{aap-14} presents two ways to strengthen these hypothesis. We focus on their large deviation approach as it is most easily verified. This approach requires identifying a \emph{rate function} $\rho:\S\times \S \to [0,\infty]$ that describes the probability of a large deviation between $F$ and $X^\varepsilon$. That is, for a sufficiently small neighborhood $U$ of a point $y$, the rate function should have the property \index{large deviations!rate function}
\[
\P[X_{t+1}^\varepsilon \in U | X_t^\varepsilon =x] \approx \exp( -\rho(x,y)/\varepsilon).
\]
Hypothesis~\ref{hy:rho} provides the precise definition and desired properties of $\rho$.

\begin{hypothesis} \label{hy:rho} There exists a \emph{rate function} $\rho: \S \times \S \rightarrow [0,+\infty]$ such that
\begin{enumerate}
\item[(i)] $\rho$ is continuous on $\S_+ \times \S$, 
\item[(ii)] $\rho(x,y) = 0$ if and only if $y = F(x)$,
\item[(iii)] for any $\beta>0$,
\begin{equation} \label{beta}
\inf \left\{ \rho(x,y) : \;  x \in \S, \, y \in \S, \, \|F(x)-y\| > \beta \right\} >0,
\end{equation} 
\item[(iv)] for any open set $U$, there is the lower bound 
\begin{equation} \label{lower}
\liminf_{\varepsilon \rightarrow 0} \varepsilon \log  p^{\varepsilon}(x,U) \geq - \inf_{y \in U} \rho(x,y)
\end{equation}
 that  holds uniformly  for $x$ in compact subsets of $\S_+$ whenever $U$ is an open ball in $\S$. Additionally, for any closed set $C$, there is the uniform upper bound
\begin{equation} \label{upper}
\limsup_{\varepsilon \rightarrow 0} \sup_{x \in \S} \varepsilon \log  p^{\varepsilon}(x,C) \leq -  \inf_{y \in C} \rho(x,y).
\end{equation}
\end{enumerate}
\end{hypothesis}
Equations \eqref{beta} and \eqref{upper}, in particular,  imply that Hypothesis \ref{H} holds.
Furthermore, as $\S_0$ is absorbing, equation \eqref{lower} implies that $\rho(x,y) = + \infty$ for all $x \in \S_0$, $y \in \S_+$. Identifying the rate function $\rho$ typically requires making use of the G\"{a}rtner-Ellis theorem~\citep[Theorem 2.3.6]{dembo-zeitouni-93} which provides large deviation estimates for sums of independent random variables. Example~\ref{return} below describes how this theorem was used for the Poisson Lotka-Volterra processes. 

We  strengthen Hypothesis~\ref{hy:M} as follows:
\begin{hypothesis} \label{hy:rho2}
For any $c>0$, there exists an open neighborhood $V_0$ of $\S_0$ such that
\begin{equation} \label{V_0}
\lim_{\varepsilon \rightarrow 0} \inf_{x \in V_0} \varepsilon \log p^{\varepsilon}(x,\S_0) \geq -c.
\end{equation}
\end{hypothesis}
\noindent Equation~\eqref{V_0} implies that 
\[
\P[X_{t+1}^\varepsilon \in \S_0 |X_t\in V_0] \ge \exp(-c/\varepsilon)
\]
for $\varepsilon>0$ sufficiently small. Namely, the probability of absorption near the boundary, at most, decays exponentially with habitat size. The following example discusses why these stronger hypotheses hold for the Poisson Lotka-Volterra process. 

\begin{example}[Return of the Poisson Lotka-Volterra Processe]\label{return}
Using the G\"{a}rtner-Ellis theorem~\citep[Theorem 2.3.6]{dembo-zeitouni-93}, \citet[Proposition 6.4]{aap-14}   showed that $\rho(x,y)=\sum_i y_i \log \frac{y_i}{F_i(x)}-y_i$ is the rate function for any Poisson processes with mean $F:\R^k_+\to\R^k_+$ including the Poisson Lotka-Volterra Process of Example~\ref{sec:LV}. To see why Hypothesis $4$ holds for the Poisson Lotka-Volterra process, assume $x$ is such that $x_i\le \delta$ for some $\delta>0$ and $i$ . Then 
\begin{eqnarray*}
\varepsilon \log \P[ X_{t+1}^\varepsilon \in \S_0 | X_t =x ]& \ge&\varepsilon \log \P[ X_{i,t+1}^\varepsilon =0| X_t =x ]\\
&=& \varepsilon \log \exp(-F_i(x)/\varepsilon) =-F_i(x)
\end{eqnarray*}
Hence, for any $c>0$, choose $\delta>0$ sufficiently small to ensure that for all $i$, $F_i(x)\le c$ whenever $x_i\le \delta$. In which case,  choosing $V_0=\{x\in \R^k_+: x_i \le \delta$ for some $i\}$ satisfies \eqref{V_0}. 
\end{example}

As many discrete distributions are used in models with demographic stochasticity (e.g. negative binomial, mixtures of bernoullis and negative binomials), an important open problem is the following:

\begin{problem}\label{a} For which types of random perturbations of an ecological model $F$ do Hypotheses 3 and 4 hold? 
\end{problem}

To relate QSDs to the attractors of the deterministic dynamics, we recall the definition of an attractor and weak* convergence of probability measures. A compact set $A\subset \S$ is an attractor for $F$ if there exists a neighborhood $U$ of $A$ such that (i) $\cap_{n\ge 1} F^n(U)=A$ and (ii) for any open set $V$ containing $A$, $F^n(U)\subset V$ for some $n\ge 1$. \index{attractor} A weak* limit point of a family of probability measures $\{\mu_\varepsilon\}_{\varepsilon>0}$ on $\S$ is a probability measure $\mu^0$ such that there exists a sequence $\varepsilon_n\downarrow 0$ satisfying \index{weak* convergence}
\[
\lim_{n\to\infty} \int h(x)\mu^{\varepsilon_n}(dx)= \int h(x) \mu^0 (dx)
\] 
for all continuous functions $h:\S\to \R$. Namely, the expectation of any continuous function with respect to $\mu^{\varepsilon_n}$ converges to its expectation with respect to $\mu^0$ as $n\to\infty$. The following theorem  follows from  \citep[Lemma 3.9 and Theorem 3.12]{aap-14}.

\begin{theorem}~\label{thm:aap1} Assume Hypotheses~\ref{hy:rho} and \ref{hy:rho2} hold. Assume for each $\varepsilon>0$, there exists a QSD $\mu_\varepsilon$ for $X^\varepsilon$. If there exists a positive attractor $A\subset \S_+$, then 
\begin{itemize}
\item there exists a neighborhood $V_0$ of $\S_0$  such that all weak* limit points $\mu^0$ of $\{\mu_\varepsilon\}_{\varepsilon>0}$ are $F$-invariant and $\mu^0(V_0)=0$, and 
\item there exists $c>0$ such that 
 \begin{equation}\label{meta}
 \lambda_\varepsilon \ge 1- e^{-c/\varepsilon} \mbox{ for all }\varepsilon>0.
 \end{equation}
 \end{itemize} Alternatively, assume that $\S_0$ is a global attractor for the dynamics of $F$. Then any weak*-limit point of $\{\mu_{\varepsilon}\}_{\varepsilon >0}$ is supported by $\S_0$.
\end{theorem}

\begin{figure}[t]
\begin{center}
\includegraphics[width=6in]{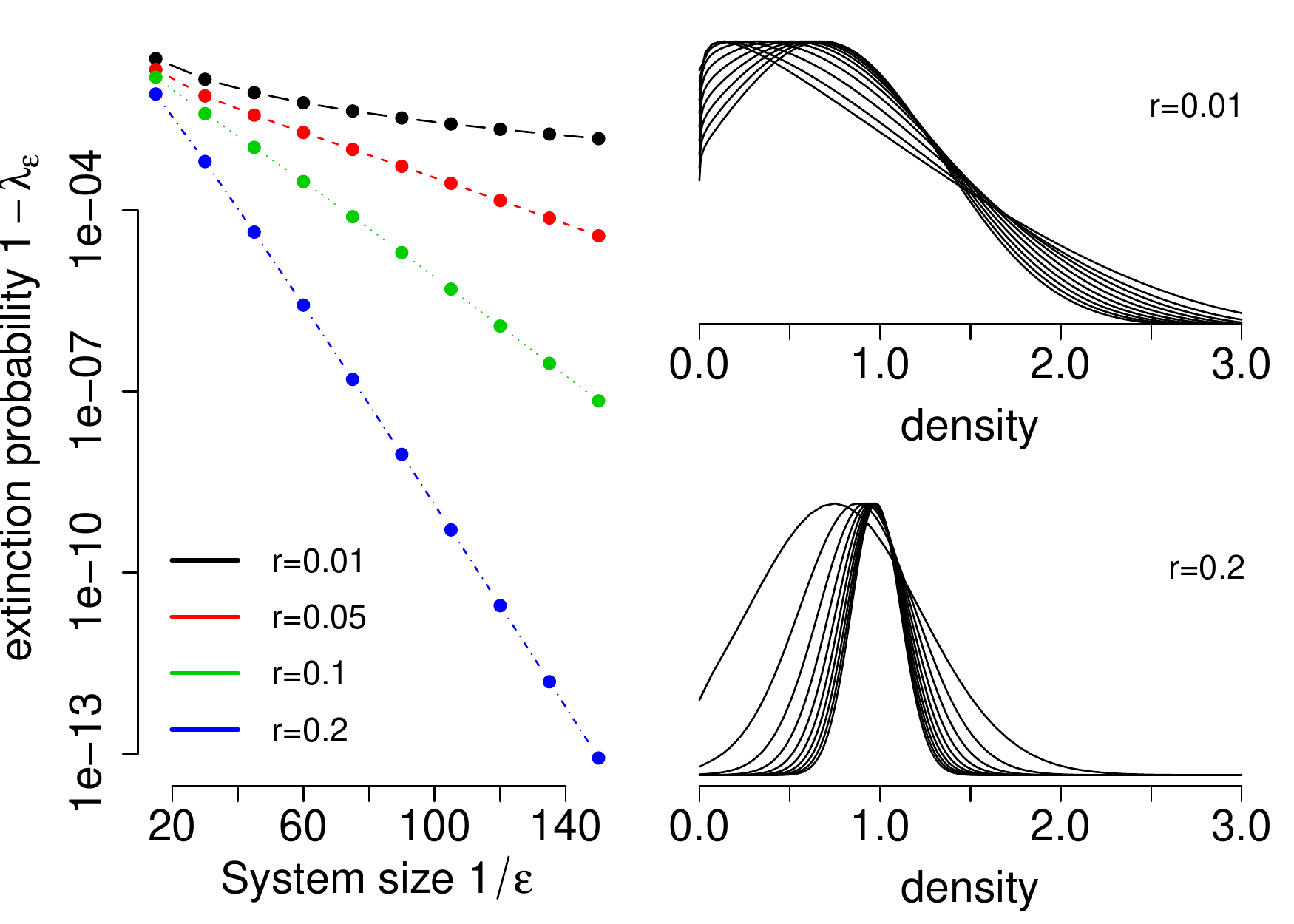}
\end{center}
\caption{Extinction probabilities and QSDs for the Poisson Ricker process described in Example~\ref{ex:Ricker}.  In the left panel, the ``intrinsic'' extinction probability $1-\lambda_\varepsilon$ plotted as function of the habitat size $1/\varepsilon$ and for different $r$ values. In the right panels, the QSDs plotted plotted for a range of habitat sizes and two $r$ values.  }\label{fig:QSD}
\end{figure}

Theorem~\ref{thm:aap1} implies the existence of a positive attractor of the deterministic dynamics ensures the stochastic process exhibits metastable behavior for large habitat size, and the probability of extinction $1-\lambda_\varepsilon$ decreases exponentially with habitat size. Equivalently, the mean time to extinction $1/(1-\lambda_\varepsilon)$ increases exponential with habitat size. These conclusions are illustrated in Figure~\ref{fig:QSD} with a one-dimensional Poisson Lotka-Volterra process (the Poisson Ricker process described below in Example~\ref{ex:Ricker}). 

Even if $F$ has  no positive attractors,  $\S_0$ may not be a global attractor as there might be an unstable invariant set in $\S_+$. For example, single species models with positive feedbacks can have an uncountable number of unstable periodic orbits despite almost every initial condition going to extinction~\citep{tpb-03}. Hence, the necessary and sufficient conditions for metastability in Theorem~\ref{thm:aap1} are not equivalent. However, if $F$ has no positive attractors, one can show that all points in $\S_+$ can with arbitrarily small perturbations be ``forced'' to $\S_0$ \citep{jtb-06,dcds-07}. Hence, this raises the following open problem.

\begin{problem}{\label{prob:extinct1} If $F$ has no positive attractors, are all the weak*-limit points of the QSDs supported by the extinction set $\S_0$?}\end{problem}

While the methodology used to prove Theorem~\ref{thm:aap1} provides an explicit expression for $c>0$, this expression is fairly abstract and only provides a fairly crude lower bound. This suggests the following questions which, if solved, may provide insights into how extinction probabilities depend on the nature of the nonlinear feedbacks within and between populations and the form of demographic stochasticity.

\begin{problem}\label{c} If $F$ has positive attractors, when does the limit 
\[
\lim_{\varepsilon\to 0} -\frac{1}{\varepsilon}\log(1- \lambda_\varepsilon)=:c
\]
exist? If the limit exists, under what circumstances can we derive explicit expressions for $c$? or good explicit lower bounds for $c$?
\end{problem}

\begin{figure}[t]
\begin{center}
\includegraphics[width=6in]{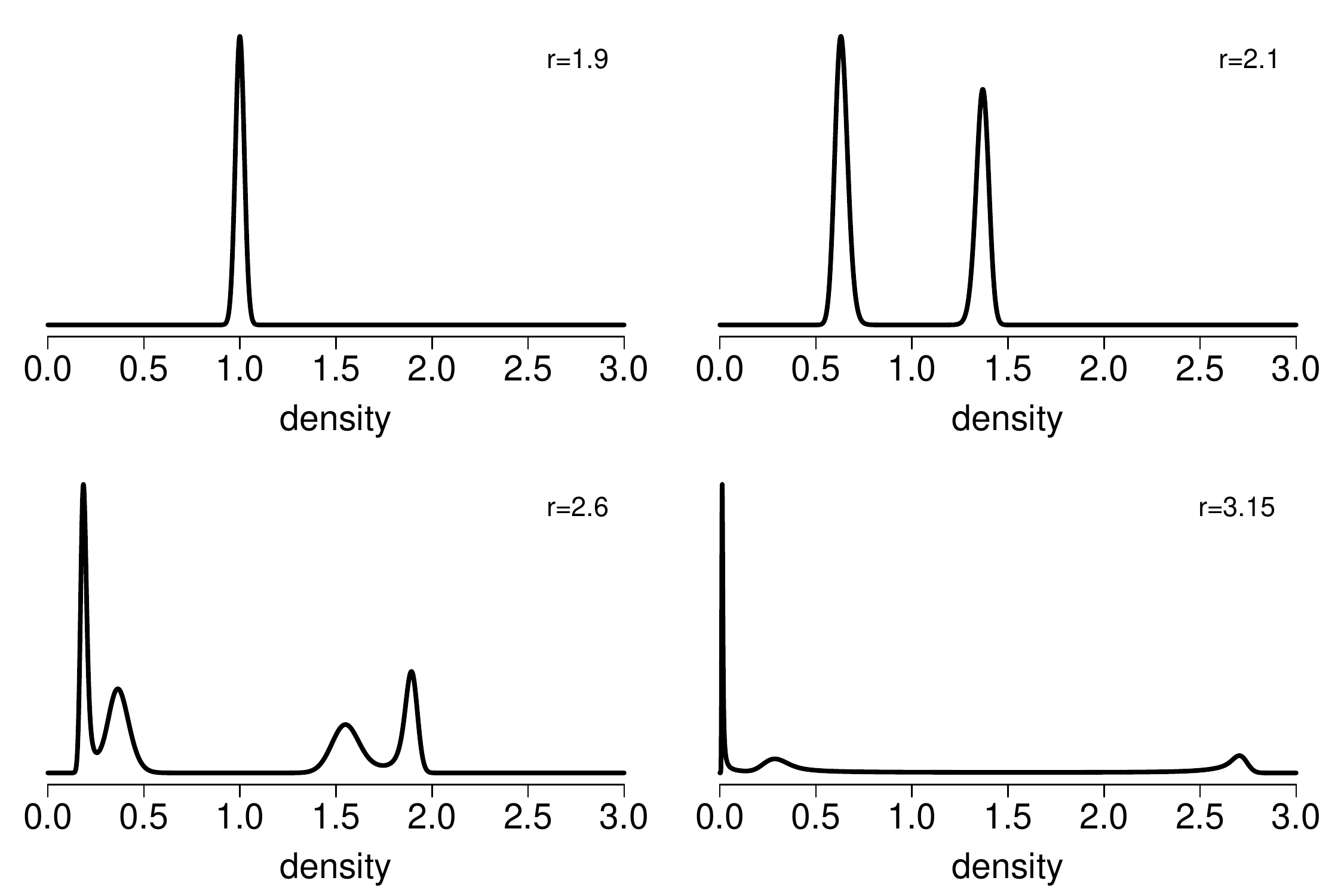}
\end{center}
\caption{QSDs for the stochastic Ricker model (see Example~\ref{ex:Ricker}) for $r$ values where $F(x)=x \exp( r (1-x))$ has a stable periodic orbit. Habitat size $1/\varepsilon$ is $2,500$. }\label{fig:Ricker}
\end{figure}

Theorem~\ref{thm:aap1} only ensures that the metastable dynamics concentrates on an invariant set for the deterministic dynamics. However, it is natural to conjecture that the QSDs $\mu_\varepsilon$ should concentrate on the positive attractors of $F$. These positive attractors, however, may coexist with complex unstable behavior. For example, the Ricker equation $F(x)=x \exp(r (1-x))$ can have a stable periodic orbit coexisting with an infinite number of unstable periodic orbits (e.g. the case of a stable period $3$ orbit as illustrated in Fig.~\ref{fig:Ricker}).

To identify when this intuition is correct, a few definitions from dynamical systems are required.  For $x\in \S$, let $\omega(x)=\{y:$ there exists $n_k\to \infty$ such that $\lim_{k\to\infty}F^{n_k}(x)=y\}$ be the \emph{$\omega$-limit set for $x$} and $\alpha(x)=\{y:$ there exist $n_k\to\infty$ and $y_k \in \S$ such that $F^{n_k}(y_k)=x$ and $\lim_{k\to\infty} y_k =y\}$ be the \emph{$\alpha$-limit set for $x$}. Our assumption that $F$ is precompact implies that there exists a global attractor given by the compact, $F$-invariant set $\Lambda=\cap_{n\ge 0} F^n(\S)$. For all $x\in \Lambda$, $\omega(x)$ and $\alpha(x)$ are compact, non-empty, $F$-invariant sets.  A \emph{Morse decomposition} of the dynamics of $F$ is  a collection of  $F$-invariant, compact sets $K_1,\dots,K_\ell$ such that 
\begin{itemize}
\item $K_i$ is isolated  i.e. there exists a neighborhood of $K_i$ such that it is the maximal $F$-invariant set in the neighborhood, and 
\item for every $x\in \Lambda\setminus \cup_{i=1}^\ell K_i$, there exist $j<i$ such that  $\alpha(x)\subset K_j$ and $\omega(x) \subset K_i$.
\end{itemize} \index{Morse decomposition}
Replacing the invariant sets $K_i$ by points, one can think of $F$ being gradient-like as all orbits move from lower indexed invariant sets to higher indexed invariant sets. Finally, recall that a compact invariant set $K$ is \emph{transitive} if there exists an $x\in K$ such that $\{x,F(x),F^2(x),\dots\}$ is dense in $K$. \citet[Theorem 2.7, Remark 2.8,  and Proposition 5.1]{aap-14} proved the following result about QSDs not concentrating on the non-attractors of $F$. The assumptions of this theorem can be verified for many ecological models. 

\begin{theorem}\label{thm:aap2} Assume Hypotheses~\ref{hy:rho} and \ref{hy:rho2} hold. Let $K_1,\dots,K_\ell$ be a Morse decomposition for $F$ such $K_j,\dots,K_\ell$ are attractors. If 
\begin{itemize}
\item $K_i \subset \S_+$ or $K_i\subset \S_0$ for each $i$,
\item $K_i\subset \S_+$ for some $ i\ge  j$, and 
\item $K_i$ with $i\le j-1$ is transitive whenever $K_i\subset \S_+$,
\end{itemize}
 then any weak*-limit point of $\{\mu_{\varepsilon}\}_{\varepsilon >0}$ is $F$-invariant and  is supported by the union of attractors in $\S_+$.
\end{theorem}

For random perturbations of deterministic models without absorbing states (e.g. models accounting for immigration or mutations between genotypes), the work of \citet{kifer-88} and \citet{freidlin-wentzell-98} can be used to show that the stationary distributions often concentrate on a unique attractor. However, due to the singularity of the rate function $\rho$ along the extinction set $\S_0$, the approach used by these authors doesn't readily extend to the stochastic models considered here. This raises the following open problem:

\begin{problem} If $F$ has multiple, positive attractors, under what conditions do the QSDs $\mu_\varepsilon$ concentrate on a unique one of these positive attractors as $\varepsilon\downarrow 0$?
\end{problem}

Lets apply some of these results to the Poisson Lotka-Volterra processes from Example~\ref{sec:LV}.

\begin{example}[The Ricker model]\label{ex:Ricker} The simplest of Poisson Lotka-Volterra processes is the stochastic Ricker model for a single species where $F(x)=x\exp(r(1-x))$ with $r>0$. \citet{kozlovski-03} proved that for an open and dense set of $r>0$ values, the Ricker map  has a Morse decomposition consisting of a finite number of unstable, intransitive sets (more specifically, hyperbolic sets) and a unique stable period orbit $\{p,F(p),\dots, F^T(p)\}$. As the stable periodic orbit is the only attractor, Theorem~\ref{thm:aap2} implies the following result. 

\begin{corollary} Consider the Ricker process with $r>0$ such that $F(x)=x\exp(r(1-x))$ has the aforementioned Morse decomposition. Then any weak*-limit point of $\{\mu_{\varepsilon}\}_{\varepsilon >0}$ is supported by the unique stable periodic orbit $\{p,F(p),\dots, F^T(p)\}$. 
\end{corollary}

Figure~\ref{fig:Ricker} illustrates this corollary: QSDs  concentrating on the stable periodic orbit of period $1$ for $r=1.9$, period $2$  for $r=2.1$, period $4$ for $r=2.6$, and period $3$ for $r=3.15$. Remarkably, in the case of the stable orbit of period $3$, there exists an infinite number of unstable periodic orbits which the QSDs do not concentrate on. We note that  \citet{hognas-97,klebaner-lazar-zeitouni-98,ramanan-zeitouni-99} proved similar results to Corollary $1$ using inherently one dimensional methods. 

\begin{figure}[t]
\includegraphics[width=0.85\textwidth]{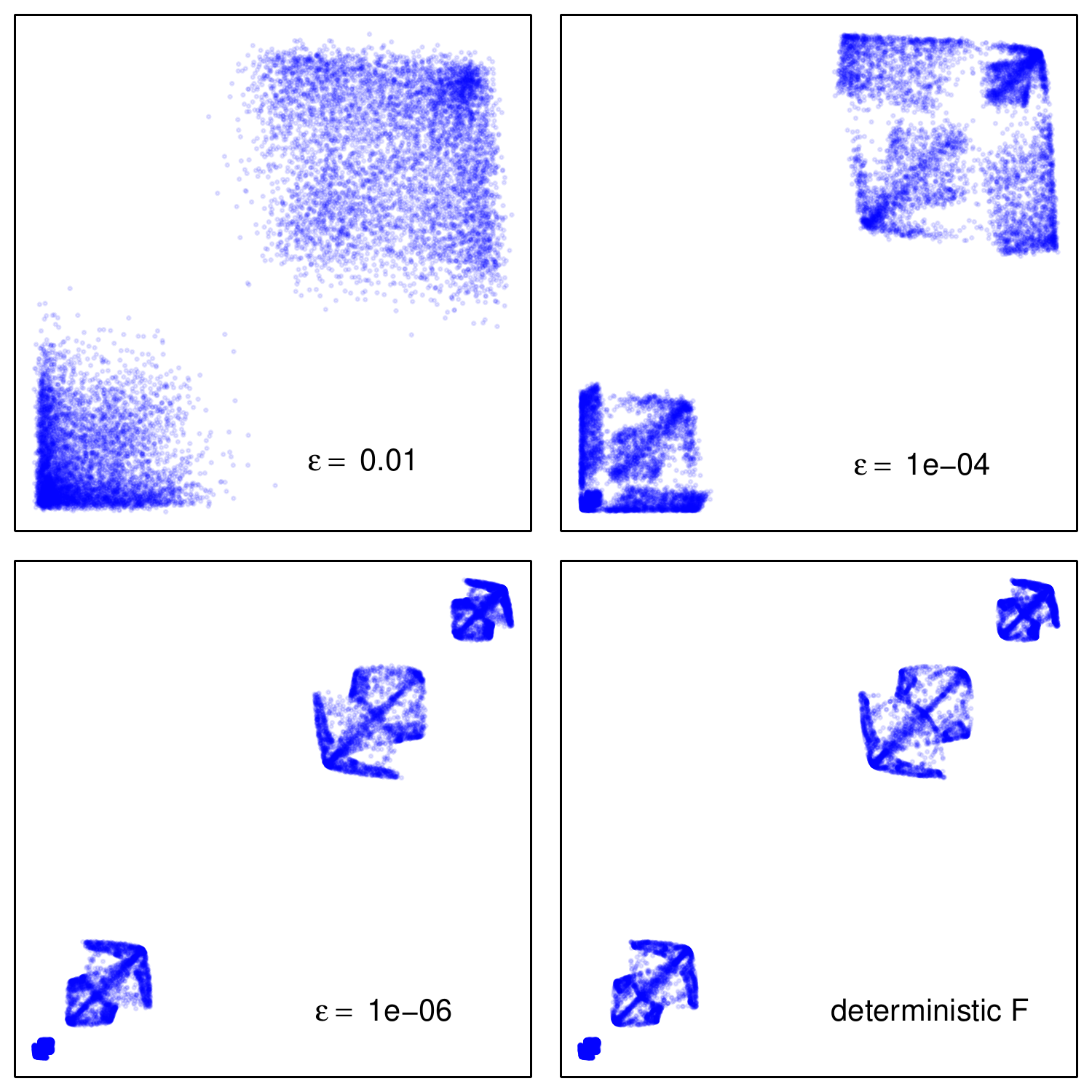}
\caption{Numerically estimated QSDs for a Poisson Lotka-Volterra process with two competing species, and the global attractor of the deterministic map $F_i(x)=x_i \exp(\sum_j A_{ij}x_jx+r_i)$. The stochastic and deterministic processes were simulated for $50,000$ time steps and the last $17,500$ time steps are plotted in the $x_1$--$x_2$ plane. Parameters:  $A$ is the matrix with rows $(-0.2,-0.01)$, $(-0.01,-0.2)$ and $r=(2.71,2.71)$.}\label{springer-2}
\end{figure}
\end{example}
\newpage
\begin{example}[Revenge of the Poisson Lotka-Volterra processes]
For higher dimensional Lotka-Volterra processes, we can use properties of Lotka-Volterra difference equations in conjunctions with Theorems~\ref{thm:aap1} and \ref{thm:aap2} to derive two algebraically verifiable results for the stochastic models. First, if the deterministic map $F=(F_1,\dots,F_k)$ with $F_i(x)=x_i  \exp(\sum_j A_{ij}x_j +r_i)$ is pre-compact and there is no internal fixed point (i.e. there is no strictly positive solution to $Ax=-r$), then \citet{hofbauer-etal-87} proved that the boundary of the positive orthant is a global attractor. Hence, Theorem~\ref{thm:aap1} implies the following corollary. 

\begin{corollary} Let $X^\varepsilon$ be a Poisson Lotka-Volterra process such that $F$ is pre-compact and admits no positive fixed point. Then any weak*-limit point of $\{\mu_{\varepsilon}\}_{\varepsilon >0}$ is supported by $\S_0$, the boundary of the positive orthant of $\R^k_+$.
\end{corollary}

On the other hand, \citet{hofbauer-etal-87} derived a simple algebraic condition which ensures that the deterministic dynamics of $F$ has a positive attractor. Namely, there exist $p_i>0$ such that
\begin{equation}\label{eq:perm}
\sum_i p_i \left(\sum_j A_{ij}x_j^*+r_i\right)>0
\end{equation}
for any fixed point $x^*$ on the boundary of the positive orthant. Hence, Theorem~\ref{thm:aap1} implies the following corollary. 

\begin{corollary} Assume $F=(F_1,\dots,F_k)$ with $F_i(x)=x_i \exp(\sum_j A_{ij}x_j+r_i)$ is pre-compact and satisfies \eqref{eq:perm} for some choice of $p_i>0$.  If $X^\varepsilon  $ is the corresponding Poisson Lotka-Volterra process, then any weak*-limit point of $\{\mu_{\varepsilon}\}_{\varepsilon >0}$ is supported by $A$ where $A\subset \S_+$ is the global, positive attractor for $F$. Moreover, there exists $c>0$ such that $\lambda_\varepsilon \ge 1-\exp(c/\varepsilon)$ for all $\varepsilon>0$ sufficiently small. 
\end{corollary}

Figure~\ref{springer-2} illustrates the convergence of the QSDs to the attractor of $F$ for a Lotka-Volterra process of two competing species. Even for populations of only hundreds of individuals ($\varepsilon=0.01$), this figure illustrates that species can coexist for tens of thousands of generations despite oscillating between low and high densities, a key signature of the underlying deterministic dynamics.  However, only at much larger habitat sizes  (e.g. $1/\varepsilon = 1,000,000$) do the metastable behaviors clearly articulate the underlying deterministic complexities. 
\end{example}

\section{Environmental stochasticity} 

To understand how environmental fluctuations, in and of themselves, influence population dynamics, we shift our attention to models for which the habitat size is sufficiently large that one can approximate the population state by a continuous variable. Specifically, let $X_t \in \R^k_+$ denote the state of the population or community at time $t$. The components of $X_t=(X_{1,t},X_{2,t},\dots,X_{k,t})$ corresponds to densities or frequencies of subpopulations.  To account for environmental fluctuations, let $\En \subset \R^m$ (for some $m$) be a compact set representing all possible environmental states e.g. all possible precipitation and temperature values. I assume that $\en_{t+1}\in \En$ represents the environmental state of the system over the time interval $(t,t+1]$ that determines how the community state changes over that time interval. If the population or community state $X_{t+1}$ depends continuously on  $\en_{t+1}$ and  $X_t$, then 
\begin{equation}\label{eq:background1}
X_{t+1}=F(X_t,\en_{t+1})
\end{equation}
for a continuous map $F:\R_+^k \times \En \to \R_+^k$. If the $\en_t$ are random variables, then \eqref{eq:background1}  is known as a \emph{continuous, random dynamical system}. \citet{arnold-98} provides a thorough overview of the general theory of these random dynamical systems. \index{stochastic difference equation}\index{random dynamical system}

To state the main hypotheses about \eqref{eq:background1}, recall that a sequence of random variables, $\en_1,\en_2,\dots,$ is \emph{stationary} if for every pair of non-negative integers $t$ and $s$, $\en_1,\dots,\en_t$ and $\en_{1+s},\dots,\en_{t+s}$ have the same distribution. The sequence is \emph{ergodic} if with probability one all realizations of the sequence have the same asymptotic statistical properties e.g. time averages (see, e.g., \citet{durrett-96} for a more precise definition). \index{stationary} \index{ergodic}

\begin{hypothesis}\label{hyp1}
 $\en_1,\en_2,\dots$ are an ergodic and stationary sequence of random variables taking value in $\En$. Let $\pi$ be the stationary distribution of this sequence i.e.  the probability measure $\pi$ on $\En$ such that $\P[\en_t\in B]=\pi(B)$ for all Borel sets $B\subset \En$.
 \end{hypothesis}
\noindent This hypothesis is satisfied for a diversity of models of environmental dynamics. For example,  $\en_t$ could be given by a finite state Markov chain on a finite number of environmental states, say $e_1,e_2,\dots, e_m\in E$ (e.g. wet and cool, wet and hot, dry and cool, dry and hot) with transition probabilities $p_{ij}=\P[\en_{t+1}=e_j|\en_t=e_i]$. If the transition matrix $P=(p_{ij})_{i,j}$ is aperiodic and irreducible, then $\en_t$ is asymptotically ergodic and stationary. Alternatively, $\en_t$ could be given by a sequence of independent and identically distributed random variables or, more generally, an autoregressive process. 

Our second hypothesis simply assumes that population densities remain bounded and allows for the possibility of extinction. 

\begin{hypothesis}\label{hyp2}
There are compact sets $\S\subset \R^k_+$ and $\S_0\subset \{x\in \S: \prod_i x_i=0\}$ such that $F:\S \times \En \to \S$, $F:\S_0 \times \En\to \S_0$, and $F:\S_+ \times \En \to \S_+$ where $\S_+=\S\setminus \S_0$.   
\end{hypothesis}

\noindent For example, $\S$ may equal $[0,M]^k$ where $M$ is the maximal density of a species or $\S$ may be  the probability simplex $\Delta=\{x\in \R^k_+: \sum_i x_i=1\}$ where $x\in \S$ corresponds to the vector of genotypic frequencies. As in the case of demographic stochasticity, $\S_0$  corresponds to the  set  where one or more populations have gone extinct. Invariance of $\S_0$ implies that once the population has gone extinct it remains extinct i.e. the ``no cats, no kittens'' principle. Invariance of $\S_+$ implies that populations can not go extinct in one time step but only asymptotically. This latter assumption is met by most (but not all) models in the population biology literature.

\begin{figure}[t]
\begin{center}
\includegraphics[width=5in]{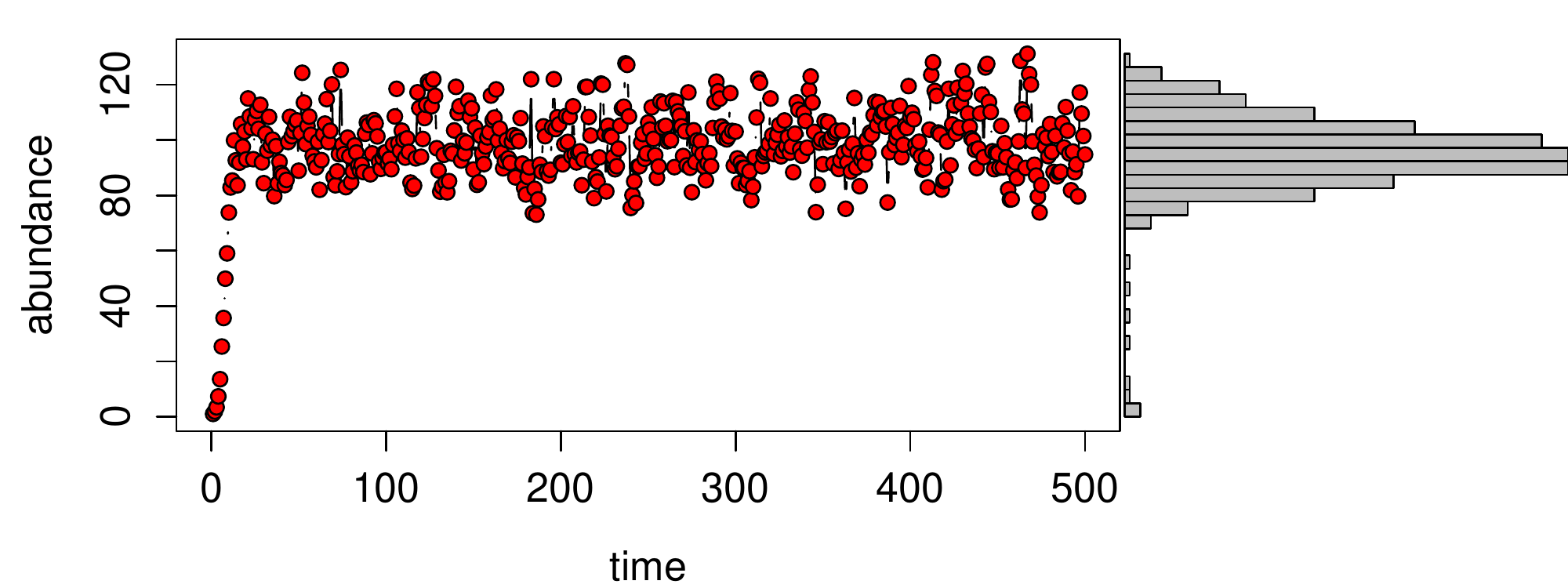}\vskip -0.4in
\includegraphics[width=3in]{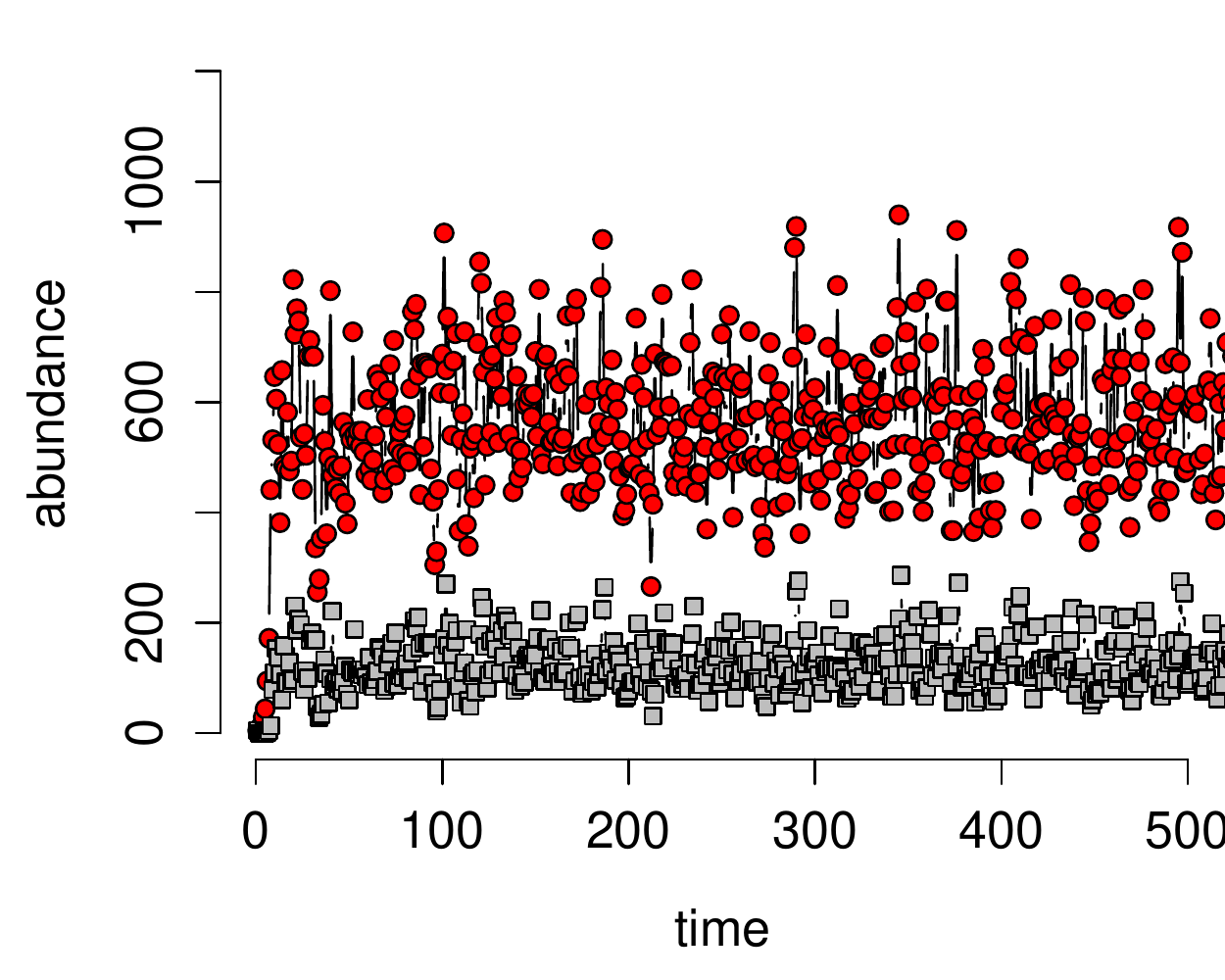}\includegraphics[width=2.6in]{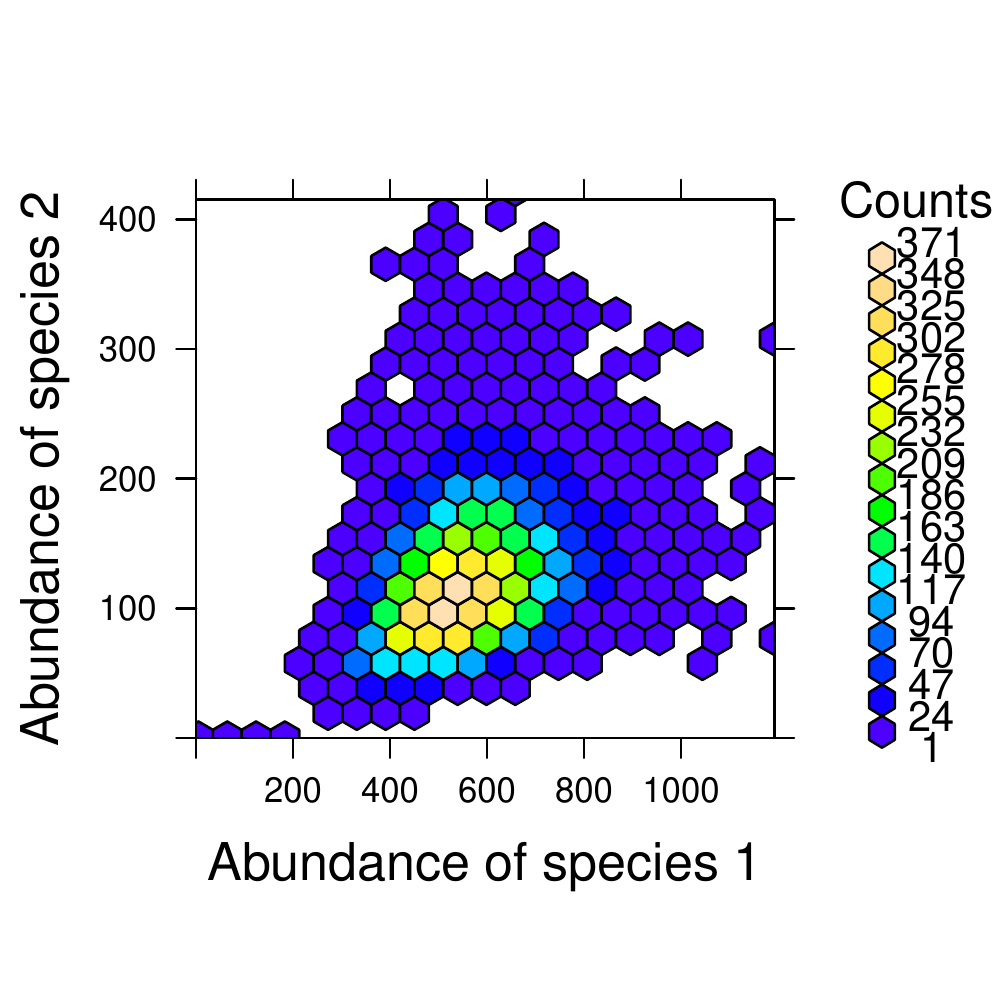}
\caption{Visualizing the empirical measures $\Pi_t^x$ for two models with environmental stochasticity. In top row, the time series of a realization of a stochastic, single species model $X_{t+1}= \frac{\eta_{t+1}X_t}{1+0.01X_t}$ where $\eta_{t}$ is a truncated log normal with log-mean $\log 2$ and log-variance $0.01$. Histogram to the right of the time series corresponds to $\Pi_{500}^x([a,b])$ for intervals $[a,b]$ of width $10$ from $0$ to $140$. In the bottom row, the time series of a realization of a stochastic predator-prey model $X_{1,t+1}=X_{1,t}\exp(\eta_{t+1}-0.001 X_{1,t}-0.001 X_{2,t})$, $X_{2,t+1}=0.5X_{1,t}(1-\exp(-0.001 X_{2,t}))$ here $\eta_{t}$ is a truncated log normal with log-mean $\log 2$ and log-variance $0.04$. To the right of the time series, the time spent in each colored hexagon in $\R_+^2$ is shown. $\Pi_{500}^x(H)$ for one of the hexagons $H\subset \R_+^2$  equals the count divided by $500$. The truncated normals are used for these models to ensure that dynamics remain in a compact set $\S$. }\label{fig:sp}\end{center}
\end{figure}

For these stochastic difference equations, there are several concepts of ``persistence'' which are reviewed in \citet{jdea-11}. Here, we focus on the ``typical trajectory'' perspective. Namely, ``how frequently does the typical population trajectory  visit a particular configuration of the population state space far into the future?''  The answer to this question is characterized by \emph{empirical measures} for  $X_t$:  \index{empirical measure}
\[
\Pi_t^x(A) = \frac{\#\{0\le s \le t: X_s\in A\}}{t+1}
\]
where $X_0=x$ and $A$ is a Borel subset of $\S$. $\Pi_t(A)$ equals the fraction of time that $X_s$ spends in the set $A$ over the time interval $[0,t]$. Provided the limit exists, the long-term frequency  that $X_t$ enters $A$ is given by $\lim_{t\to\infty} \Pi_t^x(A)$. It is important to note that these empirical measures are random measures as they depend on the particular realization of the stochastic process.  Figure 3 provides graphical illustrations of empirical measures for a single species model (top row) and a two species model (bottom row). For both models, the empirical measure at time $t$ can be approximated by a histogram describing the frequency $X_t$ spends in different parts (e.g. intervals or hexagons) of the population state space $\S$.

Stochastic persistence corresponds to the typical trajectory spending arbitrarily little time, arbitrarily near the extinction set $\S_0$. More precisely, for all $\varepsilon>0$ there exists a $\delta>0$ such that \index{stochastic persistence}\index{stochastic persistence!definition}
\[
\limsup_{t\to\infty} \Pi_t^x\left( \{x\in \S: \mbox{dist}(x,\S_0)\le \delta\}\right) \le \varepsilon\mbox{ with probability one for all } x\in \S_+
\]
where $\mbox{dist}(x,S_0)=\min_{y\in \S_0}\|x-y\|$.
In contrast to the deterministic notions of uniform persistence or permanence, stochastic persistence allows for trajectories to get arbitrarily close to extinction and only requires the frequency of these events are very small. One could insist that the trajectories never get close to extinction. However, such a definition is too strict for any model where there is a positive probability of years where the population is tending to decline e.g. the models discussed in section~\ref{sec:spatial}. Regarding this point, \citet{chesson-82} wrote
\begin{quote}
``This criterion...places restrictions on the expected frequency of fluctuations to low population levels. Given that fluctuations in the environment will continually perturb population densities, it is to be expected that any nominated population density, no matter how  small, will eventually be seen. Indeed this is the usual case in stochastic population models and is not an unreasonable postulate about the real world. Thus a reasonable persistence criterion cannot hope to do better than place restrictions on the frequencies with which such events occur.''
\end{quote}

Conditions for verifying stochastic persistence appear in papers by \citet{tpb-09,jmb-11,jdea-11,jmb-14}. As the results by \citet{jmb-14} are the most general, we focus on them. We begin with single species models and then expand to multi-species models. 

\subsection{Single species models}

Consider a single species for which an individual can be in one of  $k$  states. For example, these states may correspond to age where $k$  is the maximal age, living in one of $k$ spatial locations or ``patches'', discrete behavioral states that an individual can move between, different genotypes in an asexual population coupled by mutation, or finite number of developmental stages or size classes. $X_{i,t}$ corresponds to population density of individuals in state $i$ and $X_t=(X_{1,t},\dots,X_{k,t})$ is the population state. The population state is updated by multiplication by a $k\times k$ matrix $A(X_t,\en_{t+1})$ dependent on the population and environmental state:
\begin{equation}\label{eq:one}
X_{t+1} = A(X_t,\en_{t+1})X_t=:F(X_t,\en_{t+1}).
\end{equation}
Assume $A(X,\en)$ satisfies the following hypothesis.

\begin{hypothesis}~\label{hyp3}
$A$ is a continuous mapping from $\S \times \En$ to non-negative $k\times k$ matrices. Furthermore, there exists a non-negative, primitive matrix $B$ such that $A(x,\en)$  has the same sign structure as $B$ for all $x,\en$ i.e. the $i$--$j$-th entry of $A(x,\en)$ is positive if and only if the $i$--$j$-th entry of $B$ is positive.
\end{hypothesis} 
\noindent The primitivity assumption implies that there is a time, $T$, such that after $T$ time steps, individuals in every state contribute to individuals in all other states. Specifically, $A(X_{T-1},\en_T)A(X_{T-2},\en_{T-1})\dots A(X_0,\en_1)$ has only positive entries for any $X_0,\dots,X_{T-1}\in \S$ and $\en_1,\dots,\en_T \in \En$. This assumption is met for most models.

To determine whether or not the population has a tendency to increase or decrease when rare, we can approximate the dynamics of \eqref{eq:one} when $X_0\approx 0$ by the linearized system
\begin{equation}\label{eq:one-approx}
Z_{t+1} = B_{t+1} Z_t  \mbox{ where } Z_0=X_0 \mbox{ and }B_{t+1}=A(0,\en_{t+1}).
\end{equation}
Iterating this matrix equation gives
\[
Z_{t}=B_t B_{t-1} B_{t-2} \dots B_2 B_1 Z_0.
\]
Proposition 3.2 from \citep{ruelle-79b} and Birkhoff's ergodic theorem implies there is a quantity $r$, the dominant Lyapunov exponent, such that   \index{Lyapunov exponent!dominant}
\[
\lim_{t\to\infty}\frac{1}{t} \log \|Z_t\| = r \mbox{ with probability one}
\]
whenever $Z_0\in \R_+^k \setminus\{0\}$. Following \citet{chesson-94,chesson-00,chesson-00b}, we call $r$ the \emph{low-density per-capita growth} of the population.  When $r>0$, $Z_t$ with probability one grows exponentially and we would expect the population state $X_t$ to increase when rare. Conversely when $r<0$, $Z_t$ with probability one converges to $0$. Consistent with these predictions from the linear approximation, \citet[Theorems 3.1,5.1]{jmb-14} proved the following result. 

\begin{theorem}~\label{thm:single} Assume Hypotheses \ref{hyp1} through \ref{hyp3} hold with $\S_0=\{0\}$. 
If $r>0$, then \eqref{eq:one} is stochastically persistent. If $r<0$ and $A(0,\en)\ge A(X,\en)$ for all $X,\en$, then \[\lim_{t\to\infty} X_t =0 \mbox{ with probability one.}\]
\end{theorem}

The assumption in the partial converse is a weak form of negative-density dependence as it requires that the best conditions (in terms of magnitude of the entries of $A$) occurs at low densities.  There are cases where this might not be true e.g. models accounting for positive density-dependence, size structured models where growth to the next stage is maximal at low densities.

\begin{example}[The case of the Bay checkerspot butterflies]\label{ex:mclaughlin} The simplest case for which Theorem~\ref{thm:single} applies are unstructured models where $k=1$. In this case, $B_t=A(0,E_t)$ are scalars and \[r=\E[\log B_t].\]
The exponential $e^r$ corresponds to the geometric mean of the  $B_t$. By Jensen's inequality, the arithmetic mean $\E[B_t]$ is greater than or equal to this geometric mean $e^r$, with equality only if $B_t$ is constant with probability one. Hence, environmental fluctuations in the low-density fitnesses $B_t$ reduce $r$ and  have a detrimental effect on population persistence. \index{Bay checkerspot butterflies}

To illustrate this fundamental demographic principle, we visit a study by \citet{mclaughlin-etal-02} on the dynamics of  Bay checkerspot butterflies, a critically endangered species.  In the 1990s, two populations of this species went extinct in Northern California. The population densities for one of these populations is shown in the left hand side of Figure~\ref{fig:checkerspot1}. Both extinctions were observed to coincide with a change in precipitation variability in the 1970s (right hand side of Fig.~\ref{fig:checkerspot1}): the standard deviation in precipitation is approximately 50\% higher after 1971 than before 1971.
\begin{figure}[t]\begin{center}
\includegraphics[width=6in]{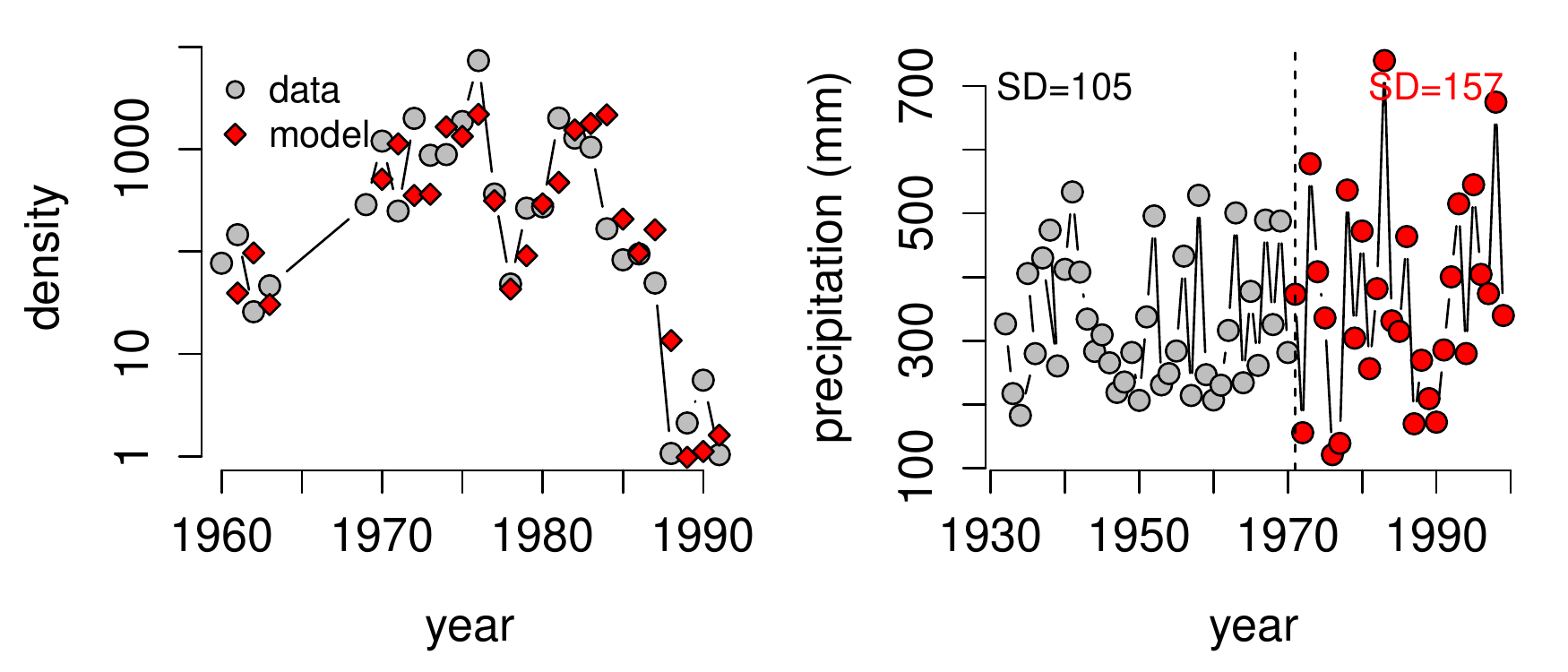} 
\end{center}\caption[Checkerspot population dynamics (left) and precipitation (right)]{Checkerspot population dynamics (left) and precipitation (right) from Example~\ref{ex:mclaughlin}. Model fit for population dynamics as red diamonds.}\label{fig:checkerspot1}
\end{figure}

To evaluate whether this shift in precipitation variability may have caused the extinction of the checkerspots, \citet{mclaughlin-etal-02} developed a stochastic difference equation of the following type 
\[
n_{t+1}=n_t\exp(a-bn_t+c\en_{t+1}^{-2})
\]
where $\en_t$ is precipitation in year $t$. Using linear regression on a log-scale yields a model whose fit for one-year predictions are shown as red diamonds in Figure~\ref{fig:checkerspot1}.
\begin{figure}[t]
\begin{center}
\includegraphics[width=6in]{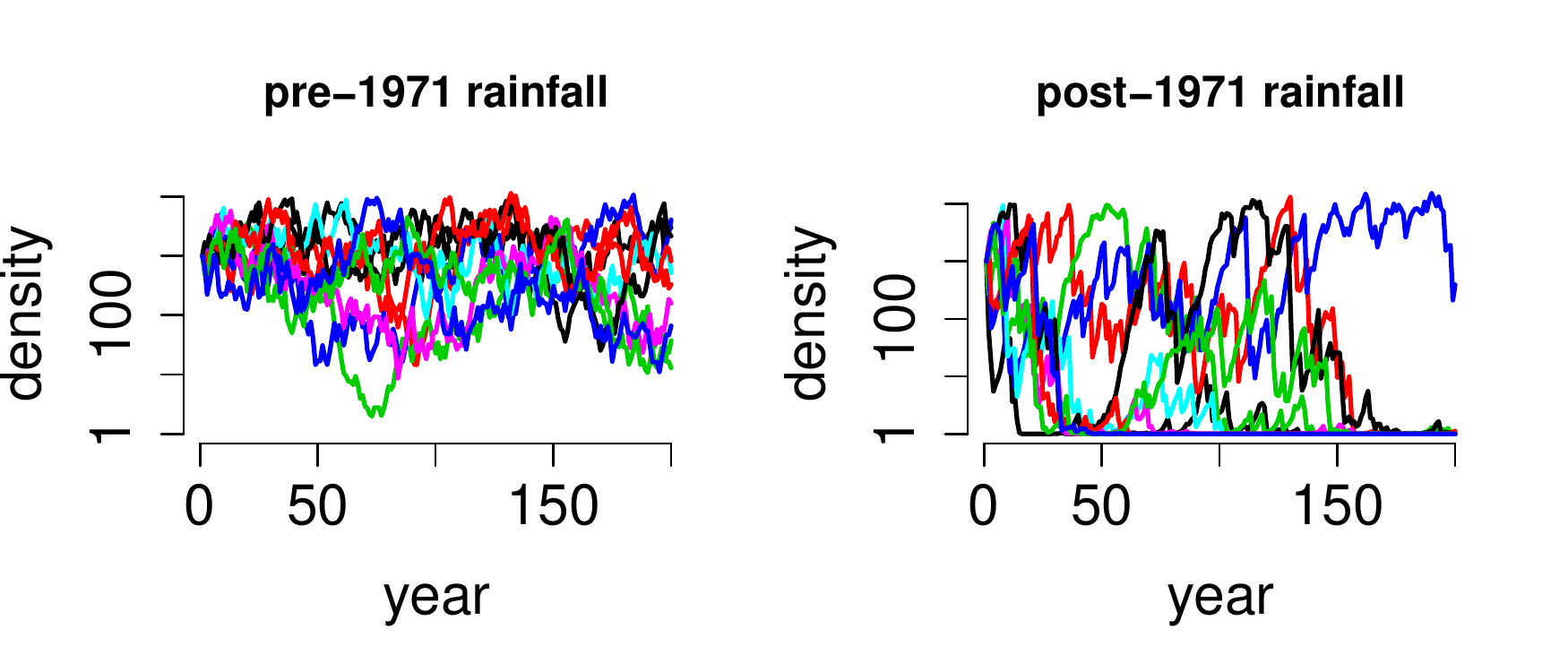}
\end{center} \caption[Simulated checkerspot population dynamics with pre-1971 precipitation data (left) and post-1971 precipitation data (right)]{Simulated checkerspot population dynamics with pre-1971 precipitation data (left) and post-1971 precipitation data (right) from Example~\ref{ex:mclaughlin}.}\label{fig:checkerspot2}
\end{figure}
To compare the pre-$1971$ and post-$1971$ population dynamics of the populations, \citet{mclaughlin-etal-02} ran their stochastic difference equations with  $\en_t$ given by independent draws from the corresponding years of precipitation data. The resulting models satisfy all of the assumptions of Theorem~\ref{thm:single}. The model with random draws from the pre-$1971$ precipitation data yields $r=\E[a+c\en_1^{-2}]=0.04$. Hence, Theorem~\ref{thm:single} implies stochastic persistence with this form of  climatic variability (left hand side of Fig.~\ref{fig:checkerspot2}). In contrast, the model with random draws from the post-$1971$ precipitation data yields $r=-0.049$. Hence, Theorem~\ref{thm:single} implies the population is extinction bound with this form of climatic variability (right hand side of Fig.~\ref{fig:checkerspot2}).
\end{example}

\begin{example}[Spatially structured populations]\label{sec:spatial} To illustrate the application of Theorem~\ref{thm:single} to structured populations, consider a population in which individuals can live in one of $k$ patches (e.g. butterflies dispersing between heath meadows, pike swimming between the northern and southern basin of a lake, acorn woodpeckers flying between canyons). $X_{i,t}$ is the population density in patch $i$. Let $E_{t+1}=(E_{1,t+1},\dots,E_{k,t+1})$ be the environmental state over $(t,t+1]$ where $E_{i,t}$ be the low-density fitness of individuals in patch $i$.  To account for within-patch competition, let $f_i(X_i,E_i)=E_i/(1+c_iX_i)$ be the fitness of an individual in patch $i$ where $c_i$ measures the strength of competition within patch $i$. This fitness function corresponds to the Beverton-Holt model in population biology. \index{spatially structured populations}

To couple the dynamics of the patches, let $d$ be the fraction of dispersing individuals that  go with equal likelihood to any other patch. In the words of Ulysses Everett McGill in  \emph{O Brother, Where Art Thou?} \begin{quote} ``Well ain't [these patches] a geographical oddity! Two weeks from everywhere!'' \end{quote} Despite this odd geographic regularity, these all-to-all coupling models have proven valuable to understanding spatial population dynamics. Under these assumptions, we get a spatially structured model of the form 
\begin{equation}\label{eq:space}
X_{i,t+1}=(1-d)f_i(X_{i,t},E_{i,t+1})X_{i,t} + \frac{d}{k-1} \sum_{j\neq i} f_j(X_{j,t},E_{j,t+1}) X_{j,t}.
\end{equation}
For this model, $A(X,E)$ is the matrix whose $i$--$j$-th entry equals $\frac{d}{k-1}f_j(X_{j,t},E_{j,t+1})$ for $j\neq i$ and $(1-d)f_i(X_{i,t},E_{i,t+1})$ for $j=i$.

The low density per-capita growth rate $r$ is the dominant Lyapunov exponent of the random product of the matrices $B_t=A(0,E_t)$. Theorem~\ref{thm:single} implies this model exhibits stochastic persistence if $r>0$ and asymptotic extinction with probability one if $r<0$. In fact, as this spatial model has some special properties (monotonicity and sublinearity), work of \citet[Theorem 1]{tpb-09} implies if $r>0$, then there is a probability measure $m$ on $\S_+$ such that 
\[
\lim_{T\to\infty} \frac{1}{T} \sum_{t=1}^T h(X_t)=\int h(x)\,m(dx) \mbox{ with probability one}
\]
for any $x\in \S_+$ and any continuous function $h:\S\to\R$. Namely, for all positive initial conditions, the long-term behavior is statistically characterized by the probability measure $m$ that places no weight on the extinction set. When this occurs, running the model once for sufficiently long describes the long-term statistical behavior for all runs with probability one. The probability measure $m$ corresponds to the marginal of an invariant measure for the stochastic model. 

But when  is $r>0$? Finding explicit, tractable formulas for $r$, in general, appears impossible. However, for  sedentary populations ($d\approx 0$) and perfectly mixing populations ($d=1-1/k$), one has explicit expressions for $r$. In the limit of $d=0$, 
\[
r=\max_i \E[\log E_{i,t}]
\]
as $f_i(0,E_i)=E_i$.  As $r$ varies continuously with $d$ (cf. \citet[Proposition 3]{tpb-09}), it follows that persistence for small $d$ (i.e. mostly sedentary populations) only occurs if $\E[\log E_{i,t}]>0$. Equivalently, the geometric mean $\exp(\E[\log E_{i,t}])$ of the low-density fitnesses $E_{i,t}$ is greater than one in at least one patch. 

When $d=1-1/k$, the fraction of individuals going from any one patch to any other patch is $1/k$. In this case, the model reduces to a scalar model for which 
\[
r=\E\left[ \log \left( \frac{1}{k}\sum_{i=1}^k E_{i,t}\right)\right].
\]
Namely, $e^r$ is equal to the geometric mean of the spatial means~\citep{metz-etal-83}. Applying Jensen's inequality to the outer and inner expressions of $r$, one gets
\[
\log \left( \frac{1}{k}\sum_{i=1}^k  \E[E_{i,t}]\right)> r > \frac{1}{k}\sum_{i=1}^k \E[\log E_{i,t}].
\]
Hence, persistence requires that the expected fitness in one patch is greater than one (i.e. $\E[E_{i,t}]>1$ for some $i$ in the left hand side), but can occur even if all the patches are unable to sustain the population (i.e. $\E[\log E_{i,t}]<0$ for all $i$ on the right hand side). Hence, local populations which are tending toward extinction (i.e. $\E[\log E_{i,t}]<0$ in all patches) can persist if they are coupled by dispersal. Even more surprising, \citet{prsb-10}
shows that stochastic persistence is possible in temporally autocorrelated environments even if $\E[E_{i,t}]<1$ for all patches. 

To better understand how $r$ depends on $d$, I make raise the following problem which has been proven have an affirmative answer for two-patch stochastic differential equation models by \citet{jmb-13}.

\begin{problem} If $E_{i,t}$ are independent and identically distributed in time and space, then is $r$ an increasing function of $d$ on the interval $(0,1-1/k)$? In particular, if $\E[\log E_{i,t}]<0<\E[\log \frac{1}{k}\sum_i E_{i,t}]$, then does there exists a $d^*\in (0,1-1/k)$ such that the population stochastically persists for $d\in (d^*,1-1/k]$ and goes asymptotically extinct with probability one for $d\in (0, d^*)$?
\end{problem}
\end{example}

\subsection{Multi-species communities}\label{sec:multi}

No species is an island as species regularly interact with other species. To account for these interactions, lets extend \eqref{eq:one} to account for $n$ species. Within species $i$, there are  $k_i$ states for individuals and $X_{i,t}=(X_{i1,t},\dots, X_{ik_i,t})$ is the vector of the densities of individuals in these different states. Then $X_t=(X_{1,t},\dots, X_{n,t})$ is the densities of all species in all of their states and corresponds to the community state at time $t$. Multiplication by a $k_i\times k_i$ matrix $A_i(X_t,\en_{t+1})$ updates the state of species $i$:
\begin{equation}\label{eq:many}
X_{i,t+1} = A_i(X_t,\en_{t+1})X_{i,t}=:F_i(X_t,\en_{t+1}) \mbox{ with }i=1,2,\dots,n.
\end{equation}
Assume that each of the $A_i$ satisfy Hypothesis~\ref{hyp3}.

To determine whether each species can increase when rare, consider the scenario where a subset of species are absent from the community (i.e. rare) and the remaining species coexist at an ergodic, stationary distribution $\mu$ for \eqref{eq:many}. Then, as in the single species case, we ask: do the rare species have a tendency to increase or decrease in this community context? Before pursuing this agenda, recall that stationarity means that $\mu$ is a probability measure on $\S\times \En$ such that (i) the marginal of $\mu$ on $\En$ is $\pi$ i.e. $\pi(B)=\mu(\S\times B)$ for all $B\subset \En$ and (ii) if $X_0,E_0$ are drawn randomly from this distribution, then $E_t,X_t$ follows this distribution for all time i.e. $\P[(X_t,E_t)\in B]=\mu(B)$ for all $t$ and Borel sets $B\subset \S\times \En$. Furthermore, ergodicity means that $\mu$ is indecomposable i.e. it can not be written as a convex combination of two other stationary distributions. Due to compactness of $\En\times \S$, stationary distributions always exist see, e.g., \citet[Theorem 1.5.8]{arnold-98}.  

By ergodicity, there exists a set of species $I\subset\{1,2,\dots,n\}$ such that $\mu$ is only supported by these species i.e. $\mu( \{x\in \S: \|x_i\|>0$ if and only if $i\in I\}\times \En)=1$. Suppose $i\notin I$ is one of the species not supported by $\mu$  and the sub-community $I$ follows the stationary dynamics i.e. $X_0,\en_0$ is randomly chosen with respect to $\mu$. To determine whether or not species $i$ has a tendency to increase or decrease when introduced at small densities $x_i=(x_{i1},\dots,x_{ik_i})\approx 0$, we can approximate the dynamics of species $i$  with the linearized system
\begin{equation}\label{eq:one-approx}
Z_{t+1} = B_{t+1} Z_t  \mbox{ where } Z_0=x_i \mbox{ and }B_{t+1}=A_i(X_t,\en_{t+1})
\end{equation}
where $X_t,\en_t$ is following the stationary distribution given by $\mu$. Iterating this matrix equation gives
\[
Z_{t}=B_t B_{t-1} B_{t-2} \dots B_2 B_1 Z_0
\]
As before, Proposition 3.2 from \citep{ruelle-79b} and Birkhoff's ergodic theorem implies there is a quantity $r_i(\mu)$ such that 
\[
\lim_{t\to\infty}\frac{1}{t} \log \|Z_t\| = r_i(\mu) \mbox{ with probability one.}
\] Lets call $r_i(\mu)$ the \emph{per-capita growth rate of species $i$ when the community is in the stationary state given by $\mu$}. For species $i\in I$ in the sub-community $I$,$r_i(\mu)$ can be defined in the same manner, but it will always equal zero~\citep[Proposition 8.19]{jmb-14}. Intuitively for species not going extinct or growing without bound, the average per-capita growth rate is zero. In the words of \citet{hardin-68}, 
\begin{quote}``a finite world can support only a finite population; therefore, population growth must eventually equal zero." \end{quote}

Using these per-capita growth rates, \citet{jmb-14} proved the following theorem. 

\begin{theorem}\label{thm:multi} Let  $\S_0=\{x \in \S: \prod \|x_i\|=0\}$.  If there exist $p_1,\dots,p_n>0$ such that 
\begin{equation}\label{eq:multi}
\sum_i p_i r_i(\mu)>0
\end{equation}
for all ergodic stationary distributions $\mu$ supported by $\S_0$,  then \eqref{eq:many} is stochastically persistent. 
\end{theorem}\index{stochastic persistence!criteria for}

The stochastic persistence condition is the stochastic analog of a condition introduced by \citet{hofbauer-81} for ordinary differential equation models. The sum in \eqref{eq:multi} is effectively only over the missing species as $r_i(\mu)=0$ for all the species supported by $\mu$. As the reverse of this condition implies that the extinction set $\S_0$ is an attractor for deterministic models, it is natural to raise the following question: 

\begin{problem}\label{prob:extinct2}
Let  $\S_0=\{x \in \S: \prod \|x_i\|=0\}$.  If there exist $p_1,\dots,p_n>0$ such that 
\[
\sum_i p_i r_i(\mu)<0
\]
for all ergodic stationary distributions $\mu$ supported by $\S_0$,
then does it follow that for all $\varepsilon>0$ there exists $\delta>0$ such that 
\[
\P[\lim_{t\to\infty} \mbox{dist}(X_t,\S_0) =0 | X_0=x]\ge 1-\varepsilon
\]
whenever $\mbox{dist}(x,\S_0)\le \delta$?
\end{problem}

For stochastic differential equations on the simplex, \citet[Theorems 4.2,5.1]{benaim-etal-08} proved affirmative answers to this problem for systems with small or large levels of noise. In their case, $\S_0$ was shown to be a global attractor with probability one. This stronger conclusion will not hold  in general. 

We illustrate Theorem~\ref{thm:multi} with applications to competing species and stochastic Lotka-Volterra differences equations. In both examples, the interacting species are unstructured i.e. $k_i=1$.

\begin{figure}[t]
\begin{center}
\includegraphics[width=3in]{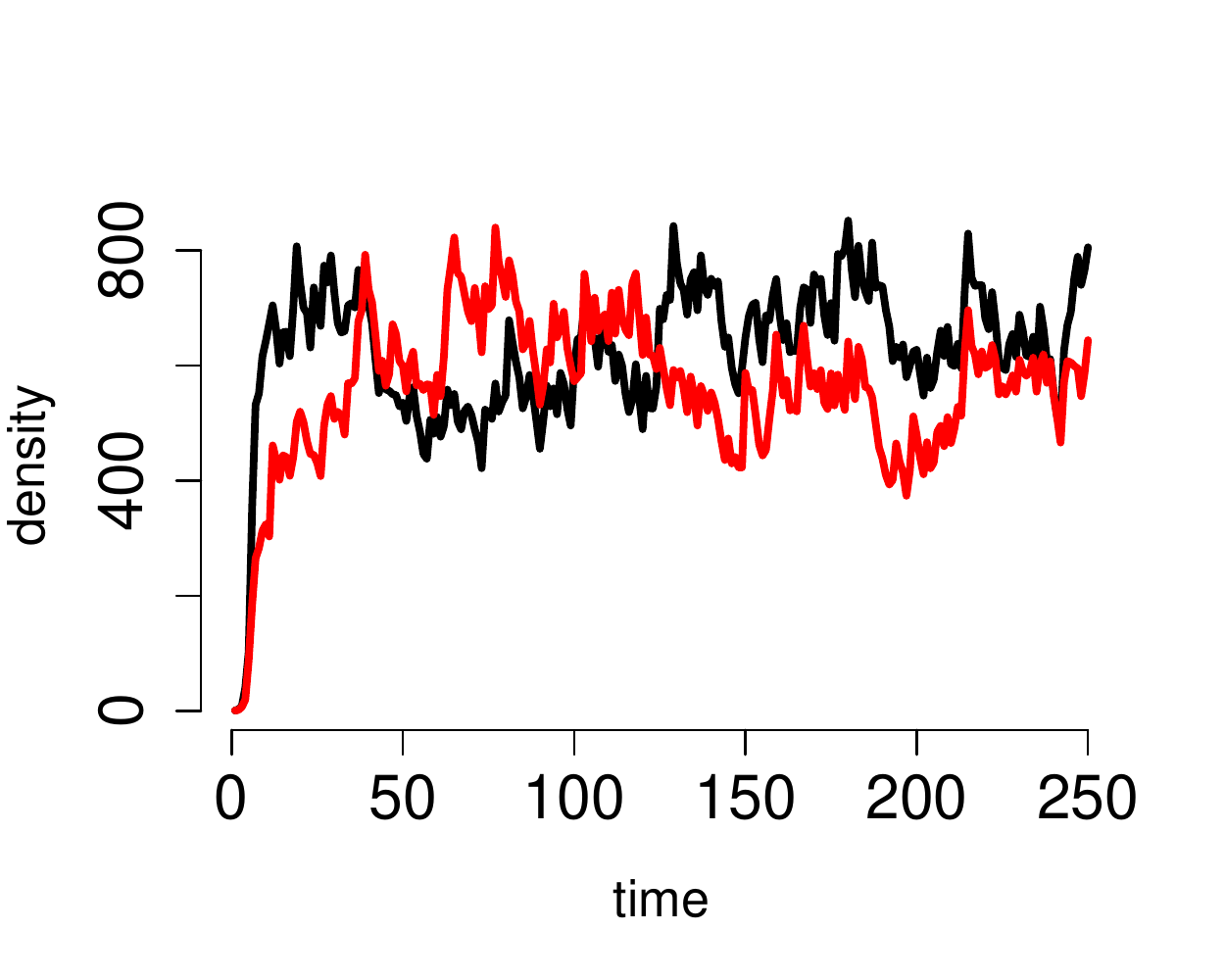}\includegraphics[width=2.6in]{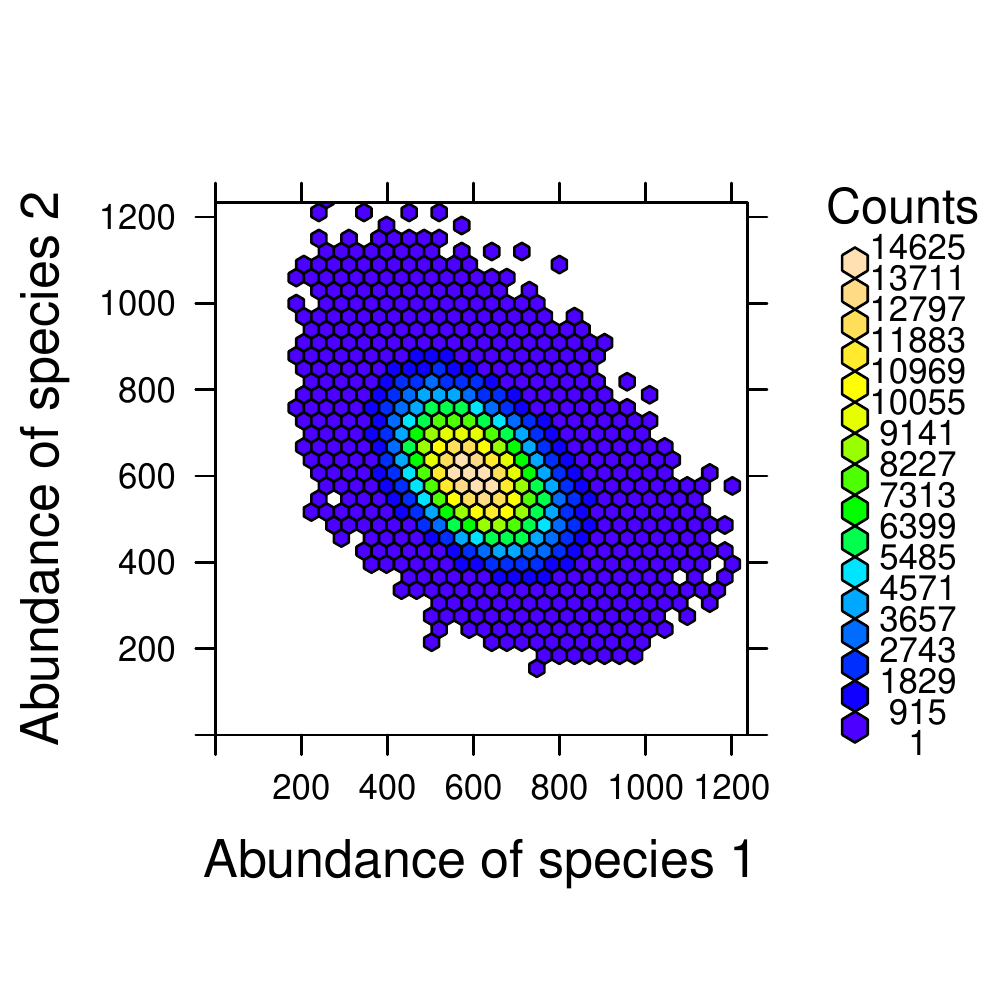}\\ \vskip -0.4in
\includegraphics[width=3in]{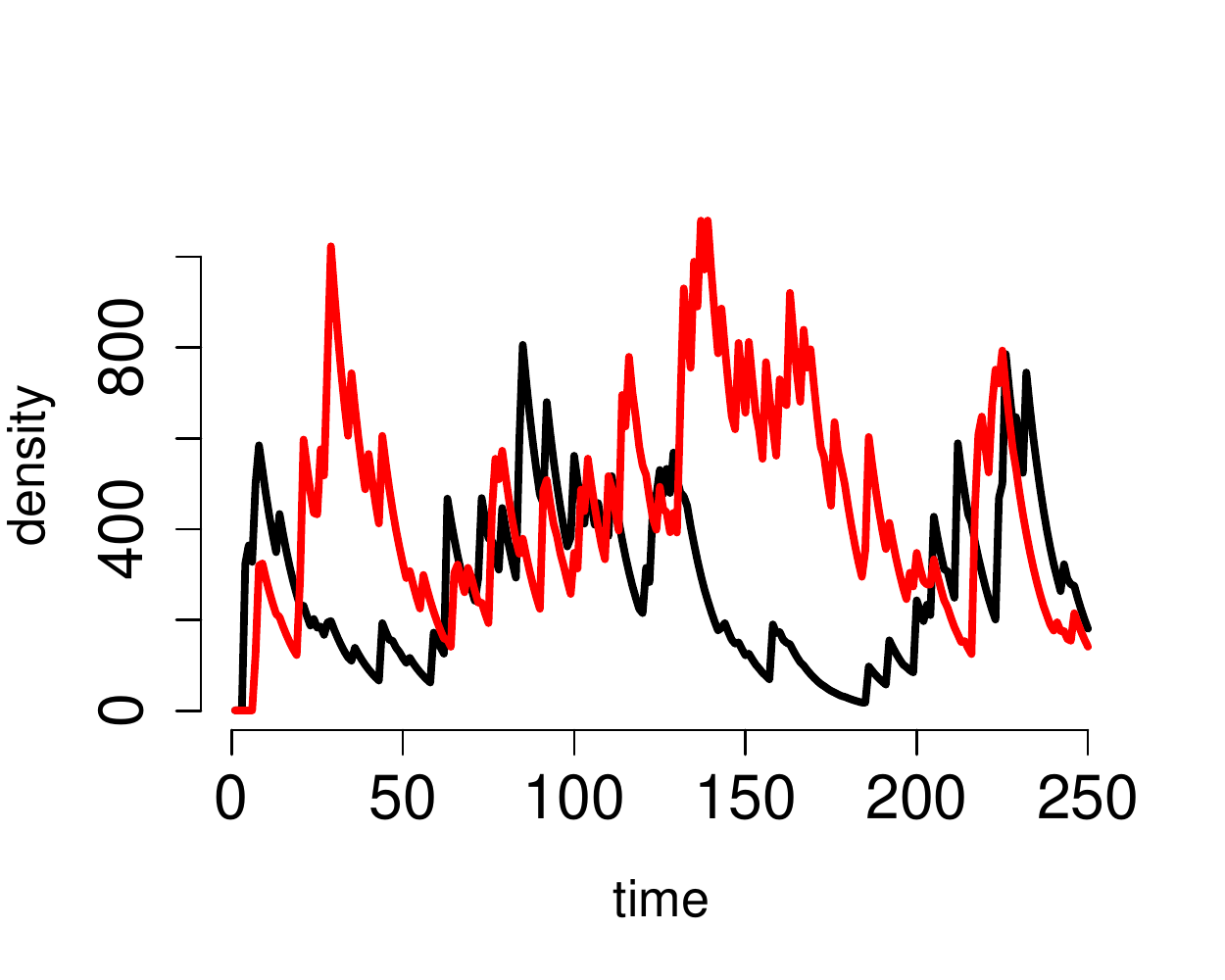}\includegraphics[width=2.6in]{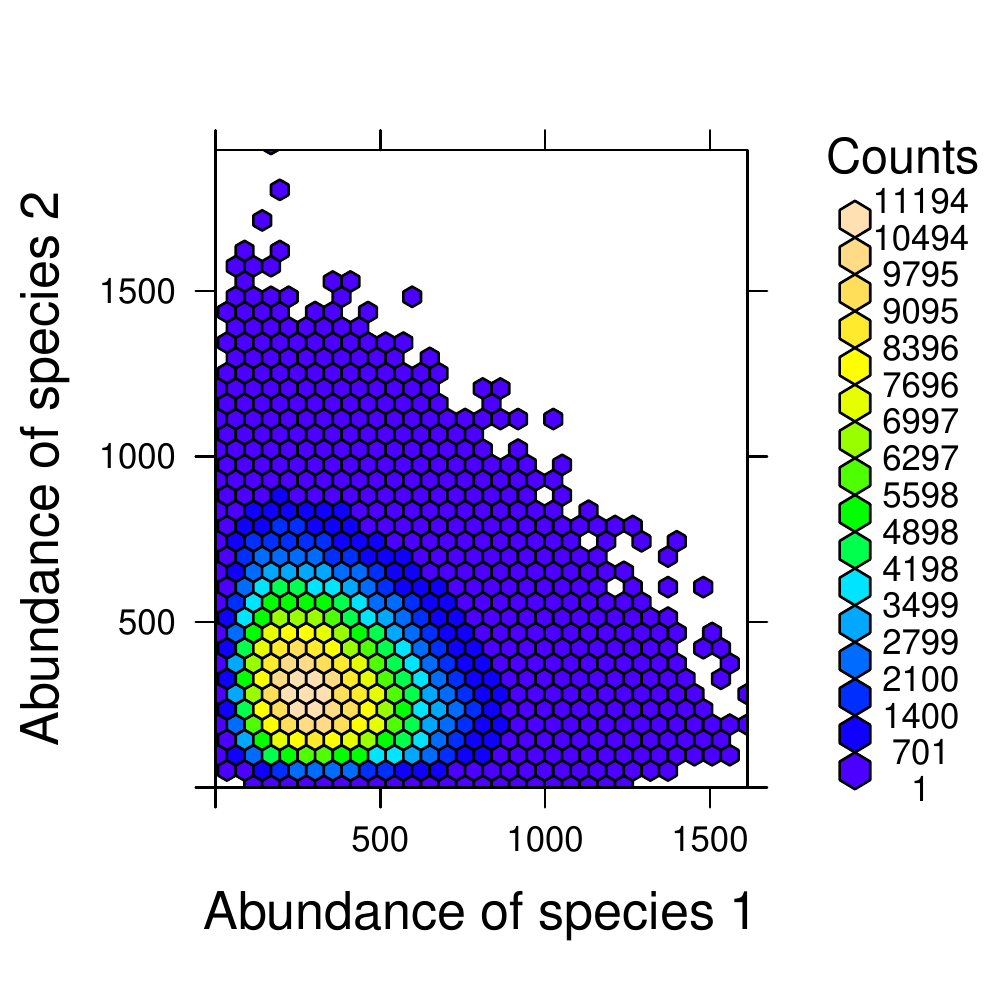}
\caption{Stochastic persistence of competing species from Example~\ref{ex:storage}. Two simulations of model \eqref{eq:comp} with $f(x)=\exp(-0.001x)$, $E_{i,t}$ truncated log normals with log-mean $1$ and log-variance $0.25$ (upper row) and $25$ (lower row). Models were run for $1,000,000$ time steps. Time series on the left show the first $250$ time steps. The two dimensional histograms on the right correspond to the last $999,000$ time steps.}\label{fig:sp}
\end{center}
\end{figure}

\begin{example}[Competing species and the storage effect]\label{ex:storage} One of the fundamental principle in ecology is the competitive exclusion principle which asserts that two species competing for a single limiting resource (e.g. space, nutrients) can not coexist at equilibrium. However, many species which appear to be competing for a single resource do coexist. One resolution to this paradox for competing planktonic species was suggested by \citet{hutchinson-61} who wrote 
\begin{quote}
``The diversity of the plankton [is] explicable primarily by a permanent failure to achieve equilibrium as the relevant external factors changes.'' 
\end{quote}
Intuitively, if environmental conditions vary such that each species has a period in which it does better than its competitors, then coexistence should be possible. Understanding exactly when this occurs is the focus of a series of papers by Peter Chesson and his collaborators~\citep{chesson-warner-81,chesson-82,chesson-88,chesson-ellner-89,chesson-94}. We illustrate one of the main conclusions from this work using a model from \citet{chesson-88}.

Consider two competing species with densities $X_t=(X_t^1,X_t^2)$ in year $t$. Let $E_{i,t}$ be the low-density per-capita reproductive output of species $i$, $s_i\in (0,1)$ the probability of adults surviving to the next year, and $f:[0,\infty)\to(0,\infty)$ a continuously differentiable, decreasing function accounting for negative effects of competition on reproduction. If $C_t=E_{1,t}X_{1,t}+E_{2,t}X_{2,t}$ represents the ``intensity of competition among the offspring'', then we have the following model of competitive interactions
\begin{equation}\label{eq:comp}
X_{i,t+1}= X_{i,t}\underbrace{\left( E_{i,t+1} f( C_t) +s_i\right)}_{A_i(X_t,E_{t+1})} \mbox{ where }C_t=E_{1,t}X_{1,t}+E_{2,t}X_{2,t}.
\end{equation}
To ensure that stochastic dynamics eventually enter a compact set $\S$, assume that $\lim_{x\to\infty}f(x)=0$ and there exists $M>0$ such that $E_{i,t}\in [0,M]$ for all $i$ and $t$. The first assumption is satisfied for many models in population biology e.g. $f(x)=\exp(-cx)$ or $\frac{1}{1+cx^b}$ with $c>0,b>0$.  

To apply Theorem~\ref{thm:multi}, we need $p_1,p_2>0$ such that $p_1r_1(\mu)+p_1r_2(\mu)>0$ for all ergodic stationary distributions $\mu$ supported by $\S_0=\{x\in S: x_1x_2=0\}$. There are three types of $\mu$ to consider: $\mu$ supports no species (i.e. $I=\emptyset$), $\mu$ only supports species $1$ (i.e. $I=\{1\}$), or $\mu$ only supports species $2$ (i.e. $I=\{2\}$). For $\mu$ supported on $\{(0,0)\}\times \En$ i.e. no species are supported, the persistence condition demands
\begin{equation}\label{c:one}
\sum_i p_i r_i(\mu)=\sum_ip_i \E[\log (E_{i,t}f(0)+s_i)]>0. 
\end{equation}
For $\mu$ supported by $\{(x_1,0):x_1>0\}\times \En$,  $r_1(\mu)=0$ and the persistence criterion requires
\begin{equation}\label{c:two}
\sum_i p_i r_i(\mu)=p_2 r_2(\mu)=p_2\int \log (E_{2}f(E_1X_1)+s_2) \mu(dXdE)>0. 
\end{equation}
As $f$ is a decreasing function, this condition being satisfied implies 
\[
\int \log (E_{2}f(0)+s_2) \mu(dXdE)=\E[\log (E_{2,t}f(0)+s_2)]>0. 
\]
Similarly, for $\mu$ supported by $\{(0,x_2):x_2>0\}\times \En$, we need
\begin{equation}\label{c:three}
\sum_i p_i r_i(\mu)=p_1 r_1(\mu)=p_1\int \log (E_{1}f(E_2X_2)+s_1) \mu(dXdE)>0. 
\end{equation}
which implies 
\[
\int \log (E_{1}f(0)+s_1) \mu(dXdE)=\E[\log (E_{1,t}f(0)+s_1)]>0. 
\]
As inequalities \eqref{c:two} and \eqref{c:three} imply inequality \eqref{c:one} for any $p_1,p_2>0$, inequalities \eqref{c:two} and \eqref{c:three} imply stochastic persistence. These inequalities correspond to the classical mutual invasibility criterion~\citep{turelli-81}: if each of the species can increase when rare, the competing species coexist.

To verify whether or not these conditions are satisfied is, in general, a challenging issue. However,  \citet{chesson-88} developed a formula for the $r_i(\mu)$ when the competition is symmetric. Namely, $s_1=s_2=:s$, $E_t$ are independent and identically distributed, and $E_{1,t},E_{2,t}$ are exchangeable i.e. $P[(E_{1,t},E_{2,t})\in B]=P[(E_{2,t},E_{1,t})\in B]$ for any Borel $B\subset \En \times \En$.  Before describing Chesson's formula, lets examine the dynamics of the deterministic case. Exchangeability and determinism imply there exists a constant $E>0$ such that $E_{1,t}=E_{2,t}=E$ for all $t$. Hence, the deterministic model is given by 
\[
x_{i,t+1}=x_{i,t} \left( E f(Ex_{1,t}+Ex_{2,t})+s\right) \mbox{ with } i=1,2.
\]
As $x_{1,t+1}/x_{2,t+1}=x_{1,t}/x_{2,t}$ for all $t$,  all radial lines in the positive orthant are invariant. Provided $Ef(0)+s>1$ (i.e. each species persists in the absence of competition), there exists a line of equilibria connecting the two axes. Regarding these neutral dynamics, \citet{chesson-88} wrote \begin{quote}``Classically, when faced with a deterministic model of this sort ecologists have concluded that only one species can persist when the likely effects a stochastic environment are taken into account. The reason for this conclusion is the argument that environmental perturbations will cause a random walk to take place in which eventually all but one species becomes extinct.''\end{quote}

Dispelling this faulty expectation, \citet{chesson-88} derived a formula for the $r_i(\mu)$. To describe this formula, assume  inequality \eqref{c:one} holds and $\mu$ is an ergodic, stationary distribution supporting species $1$. As the $E_t$ are independent in time, $\mu$ can be written as a product measure $m\times \pi$ on $\S\times \En$ where $\pi$ is given by Hypothesis~\ref{hyp1}. Define 
\[
h(E_1,E_2)=\int \log\left(E_2 f(x_1E_1)+s\right) \,m(dx).
\]
\citet{chesson-88} showed that 
\[
r_2(\mu)=-\frac{1}{2}\E\left[ \int_{E_{1,t}}^{E_{2,t}}\int_{E_{1,t}}^{E_{2,t}} \frac{\partial^2 h}{\partial E_1\partial E_2}(E_1,E_2) dE_1dE_2\right].
\]
As $f$ is a decreasing function, 
\[
 \frac{\partial^2 h}{\partial E_1\partial E_2}(E_1,E_2)=  \frac{f'(x_1E_1)x_1 s}{(E_2f(x_1E_1)+s)^2} <0
\]
whenever $s>0$. Hence, $r_2(\mu)>0$ provided that $\P[E_{1,t}>E_{2,t}]>0$ (i.e. there is some variation) and $s>0$. As this holds for any ergodic $\mu$ supporting species $1$ and a similar argument yields $r_1(\mu)>0$ for any ergodic $\mu$ supporting species $2$, it follows that this symmetric version of the model is stochastically persistent. 

The analysis of this model highlights three key ingredients required for environmental fluctuations to mediate coexistence. First, there must periods of time such that each species has a higher birth rate i.e. $E_{1,t}$ and $E_{2,t}$ vary and are not perfectly correlated. Second, year to year survivorship needs to be sufficiently positive (i.e. $s>0$ in the model) to ensure species can ``store" the gains from one favorable period to the next favorable period. Finally, the increase in fitness due to good conditions for one species is greater in years when those conditions are worse for its competitor i.e.  $\frac{\partial^2 h}{\partial E_1\partial E_2}(E_1,E_2)<0$. These are the key ingredients of the ``storage effect'' as introduced by \citet{chesson-warner-81}. \index{species coexistence!storage effect}
\end{example}

\begin{example}[Stochastic Lotka-Volterra difference equations]
Previously, we studied the Poisson Lotka-Volterra processes which injected demographic stochasticity into the discrete-time Lotka-Volterra equations~\eqref{eq:LVd}. Now, we examine the effects of injecting environmental stochasticty into these deterministic equations of $n$ interacting species:
\begin{equation}\label{LV}
X_{i,t+1}=X_{i,t} \exp\left(\sum_{j=1}^n A_{ij}X_{j,t} +b_i+E_{i,t}\right)
\end{equation}
where the matrix $A=(A_{ij})_{i,j}$ describes pairwise interactions between species, $b=(b_1,\dots,b_n)$ describes the intrinsic rates of growth of each species in the absence of environmental fluctuations, and $E_{i,t}$ describes density-independent fluctuations. \citet{turelli-81} used two dimensional versions of \eqref{LV} to examine niche overlap of competitors in random environments. 

The following lemma shows that verifying persistence for these equations reduces to a linear algebra problem. In particular, this lemma implies that the permanence criteria developed by \citet{hofbauer-etal-87} extend to these stochastically perturbed Lotka-Volterra systems. 

\begin{lemma} \label{lemmaLV}Let $\mu$ be an ergodic stationary distribution for \eqref{LV} and $I\subset\{1,\dots,k\}$ be the species supported by $\mu$ i.e.  $\mu(\{x\in \S: x^i>0$ iff $i\in I\}\times \En)=1$. Define $\beta_i = b_i+\E[E_{i,t}]$.  If there exists a unique solution $\hat x$ to
\begin{equation}\label{LV2}
\sum_{j\in I}  A_{ij}\hat x_j+\beta_i=0 \mbox{ for }i\in I\mbox{ and }
\hat x_i=0 \mbox{ for } i\notin I
\end{equation}
 then
\[
r_i(\mu)=\left\{\begin{array}{cc} 0& \mbox{ if }i\in I\\ \sum_{j\in I} A_{ij} \hat x_j +\beta_i & \mbox{otherwise.}
\end{array}\right.
\]
\end{lemma}
The following proof of this lemma is nearly identical to the proof given by \citet{jmb-11} for the case $E_t$ are independent and identically distributed in time.
\begin{proof}
Let $\mu$ and $I$ be as assumed in the statement of the lemma. We have 
\[
r_i(\mu)= \sum_{j\in I}  A_{ij} \int x_j\,\mu(dxdE)+\beta_i
\]
for all $i$. As $r_i(\mu)=0$ for all $i\in I$,
\[
0= \sum_{j\in I}  A_{ij} \int x_j\,\mu(dxdE)+\beta_i
\]
for all $i\in I$. Since we have assumed there is a unique solution $\hat x$ to this system of linear equations, it follows that $\int x_i \mu(dxdE)=\widehat x_i$ for all $i$ and the lemma follows. \qed
 \end{proof}

This lemma implies that verifying the stochastic persistence condition reduces to finding $p_1,\dots,p_n>0$ such that 
\[
\sum_{i\notin I} p_i \sum_{j\in I} A_{ij} \hat x_j +\beta_i >0
\]
for every $I \subset \{1,\dots,n\}$ and $\hat x \in \S_0$ satisfying equation \eqref{LV2}. The next example illustrates the utility of this criterion. 
\end{example}

\begin{figure}[t]
\begin{center}
\includegraphics[width=6in]{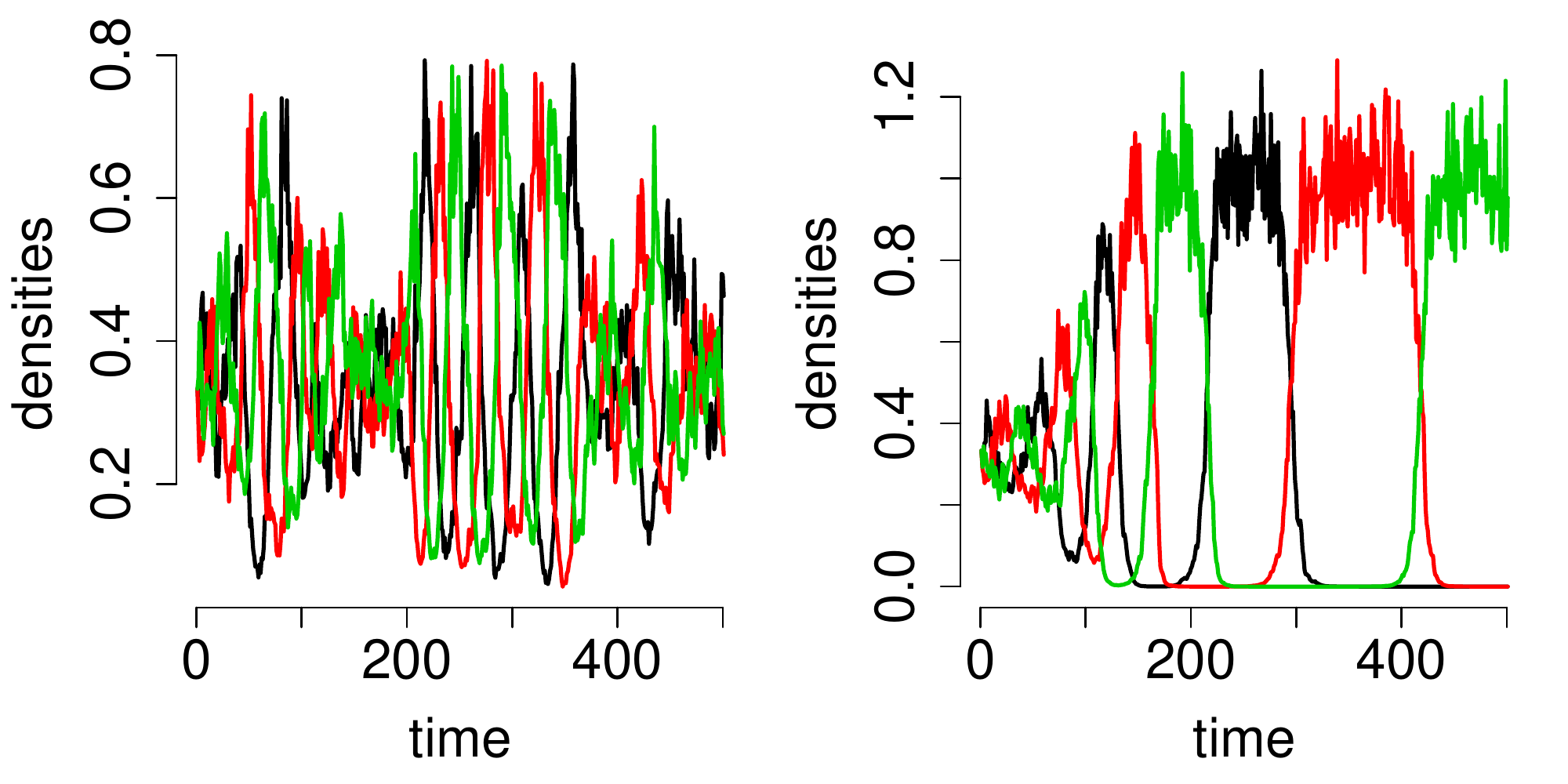}
\caption{Stochastic rock-paper-scissor dynamics (Example~\ref{ex:rps}) with stochastic persistence in the left hand panel ($w_i=0.3>0.2=\ell_i$ for all $i$) and stochastic exclusion in the right hand panel ($w_i=0.2<0.3=\ell_i$ for all $i$). }\label{fig:rps}
\end{center}
\end{figure}

\begin{example}[Rock-paper-scissor dynamics]\label{ex:rps} \index{rock-paper-scissor dynamics}\index{species coexistence!intransitive}
The Lotka-Volterra model of rock-paper-scissor dynamics is a prototype for understanding intransitive ecological outcomes~\citep{may-leonard-75,tpb-13}. Here, a simple stochastic version of this dynamic is given by \eqref{LV} with $X_1,X_2,X_3$ corresponding to the densities of the rock, paper, and scissors populations, and the matrixes $A$ and $b$ given by 
\[
A=-1+\begin{pmatrix}0& -\ell_2 &w_3 \\
w_1& 0 & -\ell_3 \\
-\ell_1& w_2& 0 
\end{pmatrix}\mbox{ and } b= \begin{pmatrix} 1\\1\\1\end{pmatrix}
\]
with $1>w_i>0$ and $\ell_i>0$. The $-\ell_i$ correspond to a reduction in the per-capita growth rate of the population losing against population $i$, and $w_i$ corresponds to the increase in the per-capita growth rate of the population winning against population $i$. Assume that the $E_{i,t}$ in \eqref{LV}  are compactly supported random variables with zero expectation. Under this assumption, $\beta_i $ as defined in Lemma~\ref{lemmaLV} equal $1$. 

Our assumptions about $A$ and $b$ imply that in pairwise interactions population $1$ is excluded by population $2$, population $2$ is excluded by population $3$, and population $3$ is excluded by population $1$. Hence, there are only four solutions of \eqref{LV2} that need to be considered: $\hat x = (0,0,0)$, $\hat x=(1,0,0)$, $\hat x= (0,1,0)$, and $\hat x=(0,0,1)$. Hence, verifying stochastic persistence reduces to determining whether there exist positive reals $p_1,p_2,p_3$ such that 
\begin{eqnarray*}
p_1 + p_2 + p_3  &>&0\\
p_1\cdot 0 + p_2 w_1 - p_3 \ell_1 &>&0\\
-p_1\ell_2 + p_2 \cdot 0  + p_3 w_2 &>&0\\
p_1w_3 - p_2 \ell_3 +p_3 \cdot 0  &>&0\\
\end{eqnarray*}
where these equation come from evaluating $\sum_i p_i r_i(\mu)$ at ergodic measures corresponding to $(0,0,0)$, $(1,0,0)$, $(0,1,0)$, and $(0,0,1)$.  Solving these linear inequalities implies that there is the desired choice of $p_i$ if and only if $w_1w_2w_3>\ell_1\ell_2\ell_3$ i.e. the geometric mean of the fitness payoffs to the winners exceeds the geometric mean of the fitness losses of the losers. Figure~\ref{fig:rps} illustrates the dynamics of coexistence when $w_1w_2w_3>\ell_1\ell_2\ell_3$  and exclusion when $w_1w_2w_3<\ell_1\ell_2\ell_3$. 
\end{example}

\section{Parting thoughts and future challenges}

The results reviewed here provide some ways to think about species coexistence or population persistence in the face of uncertainty. In the face of demographic uncertainty, species may coexist for exceptionally long periods of time prior to going extinct. I discussed how this metastable behavior may be predicted by the existence of positive attractors for the underlying deterministic dynamics, in which case the times to extinction increase exponentially with habitat size. Alternatively, in the face of environmental stochasticity, species may coexist in the sense of rarely visiting low densities. I discussed how this form of stochastic persistence can be identified by examining species' per-capita growth rates $r_i(\mu)$ when rare.  Weighted combinations of these per-capita growth rates can measure to what extent communities move away extinction as one or more species become rare.  Despite this progress, many exciting challenges lie ahead. 

Many demographic processes  and environmental conditions vary continuously in time and are better represented by continuous time models. For continuous-time Markov chains accounting for demographic stochasticity, \citet{marmet-13} proved results similar to Theorems~\ref{thm:aap1} and \ref{thm:aap2} discussed here. For stochastic differential equations of interacting, unstructured populations in fluctuating environments,  there exist some results similar to Theorem~\ref{thm:multi} by \citet{benaim-etal-08,jmb-11} and \citet{jmb-15}. These stochastic differential equations, however, fail to  account for population structure or correlated environmental fluctuations. One step toward temporally correlated environments was recently taken by \citet{benaim-lobry-14}. They characterized stochastic persistence for continuous-time models of competing species experiencing a finite number of environmental states driven by a continuous-time Markov chain. Generalizing these results to higher dimensional communities and structured populations remains an important challenge. Another exciting possibility is studying stochastic persistence for continuous-time models with stochastic birth or mortality impulses, as often observed in nature.

Biologists often measure continuous traits (e.g. body size or geographical location of an individual) that have important demographic consequences (e.g. larger individuals may produce more offspring and be more likely to survive).  Unlike models accounting for discrete traits as considered here, models with continuous traits are infinite-dimensional and, consequently, even stochastic counterparts only accounting for demographic stochasticity correspond to Markov chains on uncountable state spaces (see, e.g., \citet{mee-16}) .  One form of these models, integral projection models (IPMs), have become exceptionally popular in the ecological literature in the past decade as they interface well with demographic data sets (see, e.g., ~\citet{rees-etal-14} for a recent discussion).   Consequently, there is a need for the development of the infinite-dimensional counterparts to the results presented here  (see \citet{hardin-etal-88a} for results for structured populations facing uncorrelated, environmental stochasticity).

For both forms of stochasticity, there are few results for demonstrating that populations are ``extinction-prone'' (e.g. limiting QSDs being supported by the extinction set in Theorem~\ref{thm:aap1} or \citet[Theorems 4.2,5.1]{benaim-etal-08} for stochastic differential equations). No study of persistence or coexistence is complete without understanding this complementary outcome.  Hopefully, answers to Problems~\ref{prob:extinct1} and \ref{prob:extinct2} will narrow our gap in understanding these outcomes. Furthermore, even when populations aren't extinction prone in the aforementioned sense, extinction is inevitable as all real population are finite.  Answers to Problem~\ref{c} and their applications to specific models could provide new insights about how feedbacks between nonlinearities and noise determine the ``intrinsic'' extinction probabilities, quantities of particular importance for conservation biology.

Finally, there is the elephant in the review: what can one say for models accounting for both forms of stochasticity? At this point, all I have to offer is a natural conjecture which combines the results presented here. Namely, let $x_{t+1}=F(x_t, \en_{t+1})$ be a random difference equation and $\{X_t^\varepsilon\}_{\varepsilon>0}$ be a family of Markov chains satisfying the environmental dependent versions of Hypotheses~\ref{hy:rho} and \ref{hy:rho2} e.g. the rate function $\rho$ in Hypothesis \ref{hy:rho} depends on $\en\in\En$ as well as $x,y\in \S$.  In light of the results presented here, these models lead to the following challenging problem:
\begin{problem}
Is it true that stochastic persistence of $x_{t+1}=F(x_t,\en_{t+1})$ implies the weak* limit points of the QSDS of $\{X_t^\varepsilon\}_{\varepsilon>0}$ are supported by $\S_+$ and  $\lambda_\varepsilon\ge 1-\exp(-c/\varepsilon)$ for some $c>0$?
\end{problem}
I believe there should be an affirmative answer to this question. Namely, stochastic persistence in the face of environmental fluctuations implies long-term, persistent, metastable behavior for  communities of interacting populations of finite size, and the extinction probabilities decay exponentially with community ``size.'' Hopefully, this review will inspire work to address this problem  as well as for the other challenges  posed here. 
\vskip 0.1in
\textbf{Acknowledgments.} Many thanks to Swati Patel, William Cuello, and two anonymous reviews for providing extensive comments on an earlier version of this manuscript. This work was supported in part by US NSF Grant DMS-1313418 to the author. 

\bibliography{qsd_biblio}
\printindex
\end{document}